\documentclass[draftcls,12pt,onecolumn]{IEEEtran}

\usepackage{amssymb}
\usepackage{cite}
\usepackage{graphicx}
\usepackage{psfrag}
\usepackage{subfigure}
\usepackage{url}
\usepackage{stfloats}
\usepackage{amsmath}
\usepackage{float}
\usepackage{bm}

\usepackage{mathrsfs}
\usepackage{subeqnarray}
\usepackage{cases}
\usepackage{stfloats}
\usepackage{subeqnarray}
\usepackage{cases}

\usepackage{algorithm}
\usepackage{algorithmic}

\usepackage{xcolor}
\usepackage{soul}

\usepackage{amsthm}

\theoremstyle{plain}
\newtheorem{thm}{Theorem}
\newtheorem{lem}{Lemma}

\newtheorem{cor}{Corollary}

\theoremstyle{definition}

\theoremstyle{remark}

\definecolor{txcolor}{rgb}{0.8,0.09,0.3}

\begin{document}
%
\title{Simultaneous Wireless Information and Power Transfer in Cooperative Relay Networks with {Rateless Codes}}

\author{Xiaofei Di,
        Ke Xiong,~\IEEEmembership{Member, ~IEEE},
        Pingyi Fan,~\IEEEmembership{Senior Member, ~IEEE},\\ and Hongchuan Yang,~\IEEEmembership{Senior Member, ~IEEE} 
\thanks{
Dr. Xiaofei Di and Prof. Ke Xiong are with the School of Computer and Information Technology,  Beijing Jiaotong University,  Beijing,  P.R. China, 100044. e-mail:  \{09112084, kxiong\}@bjtu.edu.cn.

Prof. Pingyi Fan is with the Department of Electronic Engineering,  Tsinghua University, Beijing,  P.R. China, 100084. e-mail:  fpy@tsinghua.edu.cn.

Prof. Hongchuan Yang is with the Department of Electrical \& Computer Engineering, University of Victoria.  e-mail: hy@uvic.ca
}
}
\maketitle

\begin{abstract}
This paper investigates the simultaneous wireless information and power transfer (SWIPT) in cooperative relay networks, where a relay harvests energy from the radio frequency (RF) signals transmitted by a source and then uses the harvested energy to assist the information transmission from the source to its destination. {Both source and relay transmissions use rateless code, which allows the destination to employ any of the} two information receiving strategies, i.e., the mutual information accumulation (IA) and the energy accumulation (EA). {The SWIPT-enabled relay employs three different SWIPT receiver architectures, the ideal receiver and two practical receivers (i.e., the power splitting (PS) and the time switch (TS) receivers).} Accordingly, three relaying protocols, namely, ideal protocol, PS protocol and TS protocol, are presented. In order to explore the system performance limits {with these} three protocols, optimization problems are formulated to maximize their achievable information rates. For the ideal protocol, explicit expressions of the optimal solutions are derived. For the PS protocol, a linear-search algorithm is designed to solve the non-convex problems. For the TS protocol, two {solving} methods are presented.
Numerical experiments {are carried out to} validate our analysis and algorithms, which also show that, with the same SWIPT receiver, the IA-based system outperforms the EA-based system, while with the same information receiving strategy, PS protocol outperforms TS protocol. Moreover, compared with conventional non-SWIPT and non-rateless-coded systems, {the proposed protocols exhibit considerable performance gains}, especially in relatively low {signal-to-noise ratio} (SNR) regime. Besides, the effects of the source-destination direct link and the relay position on system performance are also discussed, {which {provides} insights on  SWIPT-enabled relay systems.}
\end{abstract}


%
\IEEEpeerreviewmaketitle

\section{Introduction}
Recently, integrating energy harvesting (EH) technologies into communication networks has attracted much attention \cite{Kansal}-\cite{Grover}, as it provides an effective way to implement green communications and to extend the lifetime of energy-constrained systems, including wireless sensor networks (WSN) and wireless body area networks (WBAN), etc.
In EH-enabled communication networks, EH nodes can harvest energy from surrounding environment to power their operations. Apart from the conventional {energy sources, e.g.,} solar, wind and thermoelectric energy, etc,\cite{Kansal}-\cite{Gurakan},  energy can also be harvested from radio frequency (RF) signals transmitted from other nodes with fixed energy supply such as power gird or high capacity batteries \cite{Paing}. In \cite{Varshney}, {it was} pointed out that information and energy can be simultaneously received from the RF signals as both energy and information are carried in transmitted RF signals, referred to as simultaneous wireless information and power transfer (SWIPT) \cite{Varshney}-\cite{Grover}. {Compared with conventional EH methods, SWIPT is less dependent on surrounding environments and can provide all-weather stable power supply. Therefore, it has been considered as a promising option to ensure long lifetime for energy-constrained systems.}

In preliminary works on SWIPT-enabled communications, {ideal} SWIPT receiver was assumed to be able to harvest energy and decode information from the same received signals \cite{Varshney}-\cite{Grover}. Later, this assumption was pointed out to be impractical as EH operation in the RF domain destroys the information content. Therefore, two practical receiver architectures, i.e., the power splitting (PS) receiver and the time switch (TS) receiver, were proposed in \cite{Zhang}. When PS is adopted, the received signal is split into two signal streams to perform EH and information decoding (ID), respectively. When TS is adopted, the received signal is either used to harvest energy or to decode information. Although the ideal SWIPT receiver seems {not} implementable, {it achieves an upper performance bound among all receiver architectures}  in SWIPT-enabled systems.

So far, TS and PS receiver architectures have been widely investigated in various wireless systems (see e.g. \cite{Liu}-\cite{QZhang}). In \cite{Liu}, some optimal TS policies were proposed for point-to-point communications with and without channel state information. In \cite{Huang}, \cite{Ng} and \cite{Zhou2}, they discussed {the} system throughput, energy efficiency and weighted sum-rate for {single-hop} multiuser orthogonal frequency division multiplexing (OFDM) systems, where both PS and TS receivers were considered. In \cite{QZhang}, it investigated cooperative-jamming aided robust secure transmission with PS receiver in {single-hop} multiple-input-single-output (MISO) channels and tried to maximize the worst-case secrecy rate.

Another advanced communication technology, cooperative relaying, has been widely investigated for wireless networks
{in} the past few years thanks to its advantages in  network capacity improvement, communication reliability enhancement and coverage range extension \cite{Sendonaris}-\cite{Kramer}. {That is the reason why} cooperative relaying has been listed in some communication standards, including 3GPP Release 10 and IEEE 802.16($j$).
{Relay transmission is also demonstrated as} a promising energy saving solution for wireless system by breaking a long distance transmission into several short distance transmissions \cite{Song}. {More recently,} SWIPT-enabled cooperative relay transmissions {have received much interest} as efficient solutions to prolong the life of energy-constrained wireless systems while to improve the information transmission performance, see e.g., \cite{Nasir}-\cite{Ke}.
{When} some advanced coding technologies, such as network coding and rateless coding, are employed, the performance of cooperative relaying systems can be further enhanced \cite{Katti}-\cite{Urgaonkar}. Thus, it is of significant interest to integrate advanced coding technologies into SWIPT-enabled relay systems.
However, to the best of our knowledge, only a few work discussed the SWIPT-enabled relay systems with network coding (see e.g., \cite{Moritz}-\cite{KX}) and no work has been done on rateless coded SWIPT-enabled relay systems yet.

Rateless code (RC), also termed as fountain code \cite{Luby}-\cite{Shokrollahi}, has been widely investigated to enhance the  performance of cooperative relay systems in the last decade \cite{Castura}-\cite{Urgaonkar}.  In rateless coded (RCed) systems, a transmitter generates an infinite coded stream with its original data, and the receiver is able to successfully decode and recover the original information, once its collected coded bits marginally surpass the entropy of original information. When employed in cooperative relay systems, RC allows the relay to adaptively switch from the receiving state to the transmitting state, which realizes the dynamic decode-and-forward (DF) and improves system transmission efficiency. Besides, {RC also provides an effective way to implement \emph{mutual information accumulation} (IA), which inherits the benefits of advanced information receiving methods \cite{Molisch}-\cite{Buhler}.  By accumulating mutual information from both the relay and the source, the system performance of rateless coded relaying can be greatly enhanced \cite{Molisch}-\cite{Buhler} compared with traditional relaying system with energy accumulation (EA)  information receiving, such as maximal ratio combining (MRC) method.}

Motivated by these {observations, this paper investigates the SWIPT and rateless code in a single {relay} system.}
We consider a three-node SWIPT-enabled RCed cooperative relaying network, where a source desires to transmit information to its destination with the help of a relay node. {To consider the users' battery storage limited and users' selfish nature}, {the relay may not be willing} to use its own power to help the information delivering {from} the source to the destination. By employing SWIPT technology, it is able to harvest energy from the transmitted signals from the source and then participate  {in} the  cooperative transmission. {Towards general consideration}, we assume that the source-destination direct link exists in the network. {In} such a system, both PS receiver and TS receiver are considered at the relay. For comparison, the ideal receiver is considered to offer an upper bound of the system performance. {Accordingly, three relaying protocols, namely, ideal protocol, PS protocol and TS protocol, are presented.} Besides, at the destination, both information receiving methods, IA and EA methods are considered.

Our main contributions in this paper are summarized as follows. {\textit{Firstly}, in order to explore the system performance limits of the three protocols, we formulate the corresponding } optimization problems to maximize their achievable information rates. {\emph{Secondly}, some efficient} methods are designed to solve the optimization problems on the basis of theoretical analysis. Specifically, for the ideal protocol, explicit expressions of the optimal solutions are derived. For the PS protocol, a linear search based method is designed to solve the non-convex optimization problems. For the TS protocol, the problem is first transformed into a convex problem and then is solved by using some standard convex optimization algorithms. To better understand the TS protocol, an alternative solution method is {also} provided, where the functional relationship between the system configuration parameters is provided. {\textit{Thirdly}}, numerical results are provided to validate our analysis, which show that with the same SWIPT receiver, the IA-based system outperforms the EA-based system, while with the same information receiving method, PS protocol outperforms TS protocol. Moreover, compared with conventional non-SWIPT and non rateless coded systems, by employing SWIPT and rateless code, the system performance can be greatly improved, especially at relatively low {signal-to-noise ratio} (SNR) case. In addition, the effects of the source-destination direct link and the relay position on system performance are also discussed, which provides some useful insights for better understanding and designing SWIPT-enabled RCed cooperative relaying systems.

The rest of this paper is organized as follows. Section II presents the system model. In Section III,  three relaying protocols are given and the optimization problems are formulated accordingly. Section IV proposes the corresponding analysis and algorithms to solve the optimization problems. In Section V, simulation results are provided. Finally, Section VI concludes this paper.

\section{Cooperative Relaying with Rateless Code}
\begin{figure}
  \centering
  \includegraphics[width=0.4\textwidth]{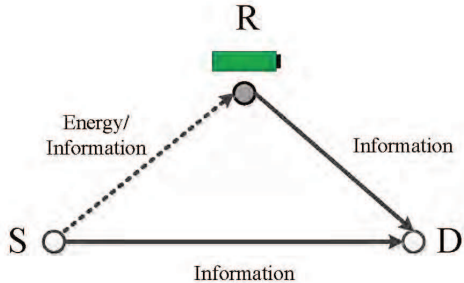}\\
  \caption{System model}\label{systemmodel}
\end{figure}

Consider a three-node relay system composed of a source \textsf{S}, a half-duplex DF relay \textsf{R}, and a destination \textsf{D}, as shown in Figure  {\ref{systemmodel}}, where all nodes are equipped with single antenna. \textsf{S} desires to transmit information to \textsf{D} with the assistance of \textsf{R}. The information is encoded with RC to transmit. It is assumed that \textsf{S} is with fixed power supply, and its available power is denoted by $P_\textsf{S}$. \textsf{R} is an energy-constrained node, which  harvests the energy from RF signals emitted by \textsf{S} and then helps the information forwarding from \textsf{S} to \textsf{D}.

All channels are modeled as quasi-static block-fading additive white Gaussian noise (AWGN) channels. The channel coefficient of the link from transmitter $u$ to its receiver $v$ is denoted by $h_{uv}$, where  $u \in \{\textsf{S},\textsf{R}\}$, $v\in \{\textsf{R},\textsf{D}\}$, and the channel power gain of link $u \rightarrow v$ is defined as $H_{uv}\triangleq|h_{uv}|^2$.
The received noise power at the \textsf{D} and \textsf{R} are denoted as $\sigma^2_\textsf{D}$ and $\sigma^2_\textsf{R}$, respectively.
In such a RCed cooperative relay SWIPT system, each cycle of transmission is divided into two phases, i.e., a broadcast phase and a collaboration phase, as illustrated in Figure  \ref{Figprotocols}. {In this paper, we adopt CDMA to realize the orthogonal transmissions over \textsf{S-D} and \textsf{R-D} links, similar to  existing works on RCed cooperative relaying (see e.g., } \cite{Molisch}-\cite{Buhler})\footnote{In fact, any of CDMA, TDMA and FDMA can be used to realize the orthogonal channel. However, if TDMA is employed, the end to end transmission delay will be increased, and if FDMA is employed, to achieve better system performance, system frequency bandwidth should be properly allocated to the R-D link and S-D link with some inter-channel guard interval. Thus, in most existing works, CDMA is adopted.}.

In the broadcast phase, \textsf{S} encodes the information with a rateless code  and then continuously broadcasts the encoded  stream to \textsf{R} and \textsf{D} with its assigned spreading codes. Once receiving the signals from \textsf{S}, \textsf{R} and \textsf{D} respectively accumulate mutual information from the information stream to try decoding the information. When the collected coded bits marginally surpass the entropy of original information, it is assumed that the receiver (\textsf{R} or \textsf{D}) can successfully decode and recover original information \cite{Buhler}. Due to the difference in link quality, \textsf{R} and \textsf{D} may decode the information in sequence.  Particularly, if \textsf{D} successfully decodes the source message before \textsf{R}, \textsf{R} will not participate in the transmission and \textsf{S} will start to {broadcast} its new messages. Whereas, if \textsf{R} successfully decodes the source message before \textsf{D}, it adaptively switches from the broadcast phase to the collaboration phase to help transmit current source message \footnote{In RCed systems, a feedback channel is often assumed, over which \textsf{R} and \textsf{D} can send a feedback to \textsf{S} in order to decide wether new messages should be transmitted or the cooperative transmission between \textsf{S} and \textsf{R} should be started, once they successfully decode the information. As only quite limited feedback information are required to be transmitted, the overhead of the feedback channel is often neglected.}.

In the collaboration phase, \textsf{R} re-encodes the message with RC and forwards the coded information to \textsf{D}. {Since CDMA is adopted to guarantee the orthogonal transmissions,} \textsf{S} can also transmit information to \textsf{D} at the same time.
Therefore, \textsf{D} may receive two information streams in the collaboration phase. One is from  \textsf{S} and the other is from \textsf{R}. If \textsf{S} and \textsf{R} adopts the same spreading codes and the same RC, \textsf{D} can use the Rake receiver to receive the signals using MRC method and perform EA-based information receiving \cite{Molisch}-\cite{Buhler}. If \textsf{S} and \textsf{R} adopts different spreading codes and independent RCs, $\textsf{D}$ is capable of separating the information from different senders (i.e., \textsf{S} and \textsf{R}) to accumulating mutual information, {i.e., the IA-based} receiving method \cite{Molisch}-\cite{Urgaonkar}. With either of IA and EA methods, when the received coded information marginally surpasses the entropy of original information, $\textsf{D}$ can successfully decode and recover original information.

Let $T$ denote the time duration needed for {\textsf{D}} to successfully decode the message and $\lambda T$ the duration {needed} for \textsf{R}. We focus on the case $0<\lambda<1$ {so that} \textsf{R} will participate the cooperative transmission. The end-to-end achievable information rate $R$ of the system is defined by $I/T$, where $I$ is the information entropy of the source message. $R$ should satisfy that \cite{Ravanshid}\footnote{The time duration of collaboration phase is $(1-\lambda )T$ and that of the broadcast phase is $\lambda T$.}
\begin{equation}\label{R}
\left\{
\begin{aligned}
&R\leq\lambda C_\textsf{SR},\\
&R\leq\lambda C_\textsf{SD}+(1-\lambda)J,
\end{aligned}
\right.
\end{equation}
where $C_\textsf{SD}$ and $C_\textsf{SR}$ represent the channel capacity of $\textsf{S}-\textsf{D}$ link and $\textsf{S}-\textsf{R}$ link, respectively. $J$ is the collaboration capacity {of the cooperative transmission in the collaboration phase}\cite{Ravanshid}, \cite{Buhler}, which depends on the adopted information receiving method.  Specifically, if IA method is adopted, $J$ can be expressed by
\begin{equation}\label{JIC}
J=\mathcal{C}\left(\frac{P_\textsf{S} H_\textsf{SD}}{\sigma^2_\textsf{D}}\right)+\mathcal{C}\left(\frac{P_\textsf{R}H_\textsf{RD}}{\sigma^2_\textsf{D}}\right),
\end{equation}
and if EA method is adopted, $J$ can be expressed by
\begin{equation}\label{JEC}
J=\mathcal{C}\left(\frac{P_\textsf{S}H_\textsf{SD}}{\sigma^2_\textsf{D}}+\frac{P_\textsf{R}H_\textsf{RD}}{\sigma^2_\textsf{D}}\right),
\end{equation}
where $\mathcal{C}(x)=\log(1+x)$, indicating that transmission rates are expressed in $nats/s$,  and $P_\textsf{R}$ denotes the transmission power of \textsf{R}, which depends on the SWIPT receiver equipped at \textsf{R}.

\section{Relaying Protocols at Energy-Harvesting Relay}\label{OPF}

\begin{figure}[t]
\centering
\includegraphics[width=0.59\textwidth]{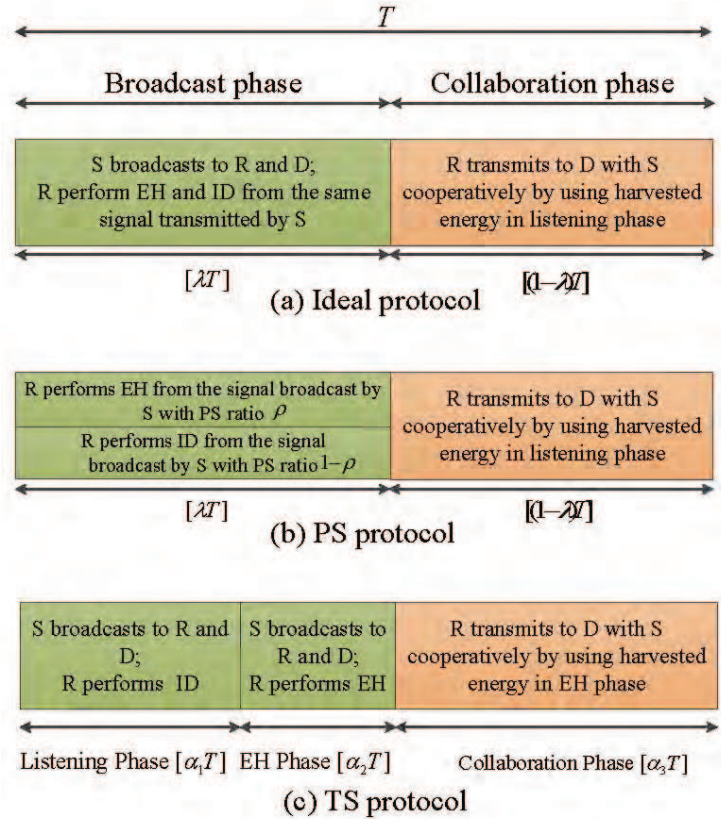}
\caption{Frameworks of the three SWIPT-enabled transmission protocols, (a) Ideal protocol; (b) PS protocol; (c) TS protocol}
\label{Figprotocols}
\end{figure}

We assume that \textsf{R} is an energy-constrained node and has to harvest energy from the RF signal transmitted from \textsf{S} to power its relaying operation. For {this} purpose, \textsf{R} must simultaneously harvest energy and decode information in the broadcast phase using a SWIPT receiver. In this work, we consider three relaying protocols, namely the ideal protocol, the PS protocol and the TS protocol, based on the adopted three SWIPT receiver architectures. The mode of operation of these protocols are illustrated in Fig. \ref{Figprotocols}. In this section, we develop the analytical expressions of $C_\textsf{SR}$ and $J$  {in (\ref{R})} for each relaying protocol.

\subsection{Ideal Protocol}
With the ideal receiver architecture, \textsf{R} is able to perform EH and ID from the same received signals, as shown in Figure  \ref{Figprotocols}(a). Thus,
the channel capacity  over $\textsf{S}\rightarrow \textsf{R}$ link can be given by
\begin{equation}
\label{CSR0}
C_\textsf{SR}^{\textrm{(Ideal)}}=\mathcal{C}\left(\frac{P_\textsf{S}H_\textsf{SR}}{\sigma^2_\textsf{R}}\right).
\end{equation}

At the same time, the energy harvested over the time period $\lambda T$ is given by \cite{Nasir}
\begin{equation}\label{E0}
E^\textrm{(Ideal)}_\textsf{R} =\lambda T\eta H_\textsf{SR}P_{\textsf{S}},
\end{equation}
where $\eta$ is a constant, describing the EH efficiency and satisfying $0\leq\eta\leq1$.
The channel capacity  over $\textsf{S}\rightarrow \textsf{D}$ link is similarly obtained as
\begin{equation}
\label{CSD0}
C_\textsf{SD}=\mathcal{C}\left(\frac{P_\textsf{S}H_\textsf{SD}}{\sigma^2_\textsf{D}}\right).
\end{equation}

Here we assume $C_\textsf{SR}^{\textrm{(Ideal)}} > C_\textsf{SD}$,  \textsf{R} can successfully decode the information before \textsf{D} and \textsf{R} is able to participate the cooperative relaying \cite{Ravanshid}-\cite{Buhler}.

Once \textsf{R} successfully decodes the information, it adaptively switches from the broadcast phase to the collaboration phase. \textsf{R} will use all harvested energy in the broadcast phase for information forwarding. Typically, the time that \textsf{D} and \textsf{R} need to successfully decodes the message, $T$ and $\lambda T${,} are random. As such, the duration of collaboration phase is also undetermined. In this work, to examine the capacity limited of the rateless-coded cooperative relay networks, we assume a block fading channel model and  that the instantaneous channel state information (CSI) is available at \textsf{S} and \textsf{R}, based on which \textsf{R} can determine the expected duration of collaboration phase $(1-\lambda)T$\footnote{{Actually, in \cite{Castura} and \cite {Ravanshid}, it was {shown} that even if the source does not know the CSI, the maximum information rate of the system also can be approached by using rateless codes, which {have} the capability to adopt to the channel state. Therefore, our analyzing methods and the obtained results also can be applied to the system without CSI known at the source.}}. As such, the transmission power in the collaboration phase at \textsf{R} is
\begin{equation}\label{PR0}
P_\textsf{R}^\textrm{(Ideal)}=\frac{E^\textrm{(Ideal)}_\textsf{R}}{(1-\lambda)T}=\frac{\eta H_\textsf{SR}P_\textsf{S}}{1/\lambda-1}.
\end{equation}

As a result, {according to (\ref{R})}, the end-to-end achievable information rate with ideal protocol can be expressed by
\begin{equation}\label{RI}
R^\textrm{(Ideal)}=\min\left\{\lambda C_\textsf{SR}^{\textrm{(Ideal)}}, \lambda C_\textsf{SD}+(1-\lambda)J^\textrm{(Ideal)}\right\},
\end{equation}
where  $J^\textrm{(Ideal)}$  is denoted as the collaboration capacity for the ideal protocol \cite{Ravanshid}, \cite{Buhler} and has different expressions for different information receiving methods. Specifically, if IA method is adopted, $J^\textrm{(Ideal)}$ can be expressed by \cite{Ravanshid}
\begin{equation}\label{JIIC}
J^\textrm{(Ideal)}=J^\textrm{(Ideal)}_{\textrm{IA}}=\mathcal{C}\left(\frac{P_\textsf{S} H_\textsf{SD}}{\sigma^2_\textsf{D}}\right)+\mathcal{C}\left(\frac{P_\textsf{R}^\textrm{(Ideal)}H_\textsf{RD}}{\sigma^2_\textsf{D}}\right),
\end{equation}
{and }if EA method is adopted, $J^\textrm{(Ideal)}$ can be expressed by \cite{Ravanshid}
\begin{equation}\label{JIEC}
J^\textrm{(Ideal)}=J^\textrm{(Ideal)}_{\textrm{EA}}=\mathcal{C}\left(\frac{P_\textsf{S}H_\textsf{SD}}{\sigma^2_\textsf{D}}+\frac{P_\textsf{R}^\textrm{(Ideal)}H_\textsf{RD}}{\sigma^2_\textsf{D}}\right).
\end{equation}


\subsection{PS Protocol}

With PS receiver architecture, \textsf{R} splits the received signals  in the broadcast phase into two streams to perform EH and ID, respectively, as shown in Figure  \ref{Figprotocols}(b). Let the power splitting ratio be $\rho$, where $0\leq\rho\leq1$ and the parts of power used for EH and ID at \textsf{R} are $\rho$ and $1-\rho$, respectively.
Thus, the channel capacity of $\textsf{S}\rightarrow \textsf{R}$ link of PS protocol is \cite{Nasir}
\begin{equation}\label{CSR}
C_\textsf{SR}^\textrm{(PS)}=\mathcal{C}\left(\frac{(1-\rho)P_\textsf{S}H_\textsf{SR}}{(1-\rho)\sigma^2_\textrm{a}+\sigma^2_\textrm{b}}\right),
\end{equation}
where $\sigma^2_\textrm{a}$ and $\sigma^2_\textrm{b}$ are the antenna noise power and signal processing noise power, respectively, and the total noise power $\sigma^2_\textsf{R}=\sigma^2_\textrm{a}+\sigma^2_\textrm{b}$.
Similar to the ideal protocol, \textsf{R} with PS protocol will perform cooperative relaying only when $C_\textsf{SR}^\textrm{(PS)}>C_\textsf{SD}$. Therefore, we have the following Lemma.

\newtheorem{lemma}{\textbf{Lemma}}
\begin{lem}
{In PS protocol, \textsf{R} will participate in the cooperative relaying of the collaboration phase only when $H_\textsf{SR}>H_\textsf{SD}$ and $\rho$  satisfies}
\begin{equation}\label{PSrho}
0\leq\rho<\rho^{\textrm{th}},
\end{equation}
where $\rho^{\textrm{th}}=1-\frac{{H_\textsf{SD}\sigma^2_\textrm{b}}}{H_\textsf{SR}{\sigma^2_\textsf{D}}-{H_\textsf{SD}\sigma^2_\textrm{a}}}$. \label{T3}
\end{lem}
\begin{IEEEproof}
From (\ref{CSR}), it can be easily proved that $C_\textsf{SR}^\textrm{(PS)}$ is a decreasing function of $\rho$, and when $\rho=0$, $C_\textsf{SR}^\textrm{(PS)}$ achieves its maximum, {which} equals to $C_\textsf{SR}^{\textrm{(Ideal)}}$. When $H_\textsf{SR}\leq H_\textsf{SD}$, $C_\textsf{SR}^\textrm{(PS)}$ is always smaller than $C_\textsf{SD}$. Thus, it will not participate in the cooperative relaying. Whereas, when $H_\textsf{SR}\geq {H_\textsf{SD}}$, $C_\textsf{SR}^\textrm{(PS)}$ is possible to hold. To guarantee $C_\textsf{SR}^\textrm{(PS)}$, $\rho$ should satisfy that
$
0\leq\rho<\rho^{\textrm{th}},
$
where $\rho^{\textrm{th}}$ is obtained by solving $C_\textsf{SR}^\textrm{(PS)}=C_\textsf{SD}$ and its value is  $\rho^{\textrm{th}}=1-\frac{{H_\textsf{SD}\sigma^2_\textrm{b}}}{H_\textsf{SR}{\sigma^2_\textsf{D}}-{H_\textsf{SD}\sigma^2_\textrm{a}}}$.
\end{IEEEproof}

Here we assume that $\rho$ is always less than $\rho^{\textrm{th}}$.  Note that the energy harvested at \textsf{R} in the broadcast phase now becomes \cite{Nasir}
\begin{equation}\label{E}
E^\textrm{(PS)}_\textsf{R} =\lambda T\eta {\rho}H_\textsf{SR}P_{\textsf{S}}.
\end{equation}

With the knowledge of the instantaneous channel state information, the available power for information forwarding in the collaboration phase at \textsf{R} can be determined as
\begin{equation}\label{PR}
P_\textsf{R}^\textrm{(PS)}=\frac{E^\textrm{(PS)}_\textsf{R}}{(1-\lambda)T}=\frac{\eta\rho H_\textsf{SR}P_\textsf{S}}{1/\lambda-1}.
\end{equation}

If we denote $J^\textrm{(PS)}$ as the collaboration capacity associated with the collaboration phase of PS protocol, the end-to-end achievable rate of the system with PS relaying can be given by
\begin{equation}\label{RCT}
R^\textrm{(PS)}=\min\{\lambda C_\textsf{SR}^\textrm{(PS)}, \lambda C_\textsf{SD}+(1-\lambda)J^\textrm{(PS)}\},
\end{equation}
{where the expressions of $J^\textrm{(PS)}$ associated with IA and EA method can be obtained by substituting (\ref{PR}) into (\ref{JIC}) and (\ref{JEC}), respectively.}



\subsection{TS Protocol}
For TS protocol, the broadcast phase is further divided into two sub-phases, i.e., the listening sub-phase and the EH sub-phase, as shown in Figure  \ref{Figprotocols}(c). In the listening sub-phase, \textsf{R} receives signals from \textsf{S} and decodes the information, and in the EH sub-phase, it receives signals from \textsf{S} and harvests energy. In the collaboration phase, \textsf{S} and \textsf{R} cooperatively transmit information to \textsf{D}, which is the same with the corresponding operations of the ideal and the PS protocols. For clarity, the time fractions of the listening sub-phase, the energy harvesting sub-phase and the collaboration phase are denoted as $\alpha_1$, $\alpha_2$ and $\alpha_3$, respectively, where $\alpha_1, \alpha_2, \alpha_3\geq0$ and $\alpha_1+\alpha_2+\alpha_3=1$.

The achievable information rate over $\textsf{S}\rightarrow \textsf{R}$ link for TS protocol is
{\begin{equation}\label{CSRT}
C_\textsf{SR}^{\textrm{(TS)}}=\mathcal{C}\left(\frac{P_\textsf{S}H_\textsf{SR}}{\sigma^2_\textsf{R}}\right).
\end{equation}}
It can be seen that $C_\textsf{SR}^\textrm{(TS)}$ is the same with $C_\textsf{SR}^{\textrm{(Ideal)}}$. So, similar to the ideal protocol, in TS protocol, if the collaboration phase is involved, it also requires that ${C_\textsf{SR}^{\textrm{(TS)}}}>{C_\textsf{SD}}$, i.e., {$H_\textsf{SR}\geq H_\textsf{SD}$}, which guarantees \textsf{R} successfully decoding information before \textsf{D}.
Within $\alpha_2T$, the energy harvested at \textsf{R} in the EH sub-phase is \cite{Nasir}
\begin{equation}\label{ET}
E^\textrm{(TS)}_\textsf{R} =\alpha_2 T\eta H_\textsf{SR}P_{\textsf{S}}.
\end{equation}

{Moreover, as all harvested information is used for the information transmission,} the available power in the collaboration phase at \textsf{R} can be given by
\begin{equation}\label{PRT}
P_\textsf{R}^\textrm{(TS)}=\frac{E^\textrm{(TS)}_\textsf{R}}{\alpha_3T}=\frac{\alpha_2\eta H_\textsf{SR}P_\textsf{S}}{\alpha_3}.
\end{equation}

If we denote $J^\textrm{(TS)}$ as the collaboration capacity for TS protocol, the end-to-end achievable rate of the system for TS protocol can be given by
\begin{equation}\label{RT}
R^\textrm{(TS)}=\min\{\alpha_1 C_\textsf{SR}^\textrm{(TS)}, (\alpha_1+\alpha_2) C_\textsf{SD}+\alpha_3J^\textrm{(TS)}\},
\end{equation}
{where the expressions of $J^\textrm{(TS)}$ associated with IA and EA method can be obtained by substituting (\ref{PRT}) into (\ref{JIC}) and (\ref{JEC}), respectively.}

\section{Optimal Design for Achievable Rate Maximization}

To explore the performance limits of the three relaying protocols for EH relaying with rateless code, we formulate several optimization {problems} to maximize the achievable information rate of the system. In particular, we optimize the values of transmission parameters, including $\lambda$, $\rho$, and $\alpha_i$ ($i=1,2,3$) based on the end-to-end achievable rate expression derived in previous section. Note that with the knowledge of the instantaneous channel state information of all links, \textsf{S} may adjust the amount of information entropy in the transmitted message to achieve a certain desired $\lambda$ value. Then, $\rho$ and  $\alpha_i$ can be similarly determined.
For convenience and clarity, we define some notations used in the analysis at first, where $a\triangleq C_\textsf{SR}^{\textrm{(Ideal)}}-C_\textsf{SD}$, $b\triangleq C_\textsf{SD}$, $c\triangleq \frac{\eta H_\textsf{SR}H_\textsf{RD}P_\textsf{S}}{\sigma^2_\textsf{D}}$, and $m\triangleq\frac{P_\textsf{S}H_\textsf{SD}}{\sigma^2_\textsf{D}}$.
\subsection{Ideal Protocol}
In this subsection, we consider the optimal design of the ideal protocol. Specifically, we formulate the following uniform optimization problem to maximize the achievable information rate of the system {as}
\begin{flalign}
\textbf{P1}:\,\,\max\limits_{\lambda}\,\,&R^\textrm{(Ideal)}=\min\left\{\lambda C_\textsf{SR}^{\textrm{(Ideal)}}, \lambda C_\textsf{SD}+(1-\lambda)J^\textrm{(Ideal)}\right\}\\
\textrm{s.t.}\,\,&0<\lambda<1.\nonumber
\end{flalign}
We now consider the solution of this problem for IA-based ideal protocol and EA-based ideal protocol, separately.
\subsubsection{{Ideal protocol with IA method}}

We first analyze the case that IA method is adopted, for which we obtain the following lemma.
\begin{lem}
{Let $f_1(\lambda)=\lambda C_\textsf{SD}+(1-\lambda)J^\textrm{(Ideal)}_{\textrm{IA}}$. $f_1(\lambda)$ is a concave function w.r.t variable $\lambda$ for $\lambda \in (0,1)$.}
\label{T1}
\end{lem}
\begin{IEEEproof}
The {first-order} derivative of $f_1(\lambda)$ w.r.t $\lambda$ is
$
f_1'(\lambda)=-\log\left(1+\frac{c\lambda}{1-\lambda}\right)+\frac{(c-1)(1-\lambda)}{1+(c-1)\lambda}+1,
$
and the {second-order} derivative is
$
f_1''(\lambda)=-\frac{c^2}{(1+(c-1)\lambda)^2(1-\lambda)}.
$
When $\lambda \in (0,1)$, $f_1''(\lambda)\leq0$ always holds, so $f_1(\lambda)$ is a concave function of $\lambda$.
\end{IEEEproof}

{Note that $c\geq 0$.} It can be seen that when $c=0$, $f_1(\lambda)$ is a constant function of $\lambda$, i.e., $f_1(\lambda)=C_\textsf{SD}$. Then (\ref{RI}) is {reduced} to $R^\textrm{(Ideal)}=\min\{\lambda C_\textsf{SR}^{\textrm{(Ideal)}}, C_\textsf{SD}\}$. In this case, one can easily obtain the solution of Problem \textbf{P1} and the corresponding optimal $\lambda^*$ can be any value in the interval $[C_\textsf{SD}/C_\textsf{SR}^{\textrm{(Ideal)}},1)$. But for $c>0$ case, it becomes more complex and we discuss its optimal solution as follows. In the $c>0$ case, since $f_1(\lambda)$ is a strictly concave function of $\lambda$, we can obtain the following theorem.

\begin{figure}[t]
\centering
\includegraphics[width=0.49\textwidth]{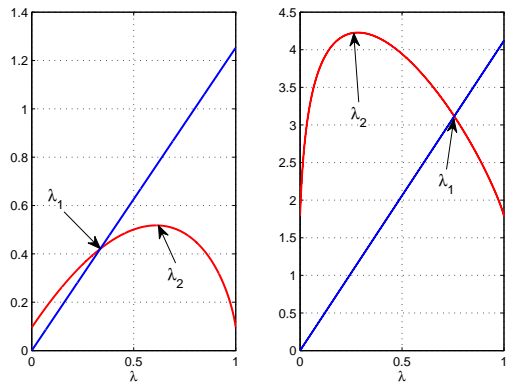}
\caption{The figure plotting $f_1(\lambda)$ (red curve) and $f_3(\lambda)=\lambda C_\textsf{SR}^{\textrm{(Ideal)}}$ (blue curve) vs. $\lambda$. The left and right subfigures represent two cases of the intersections of two curves. $\lambda_1$ is the $\lambda$-coordinate of the intersection, and $\lambda_2$ is the $\lambda$-coordinate of  the stagnation point of $f_1(\lambda)$.}
\label{Fig2}
\end{figure}

\newtheorem{theorem}{\textbf{Theorem}}
\begin{thm}
\label{T2}
The optimal solution $\lambda^*$ of IA method of Problem \textbf{P1} to obtain the maximum achievable rate can be calculated by
$\lambda^*=\max\{\lambda_1,\lambda_2\}$,
where $\lambda_1$ is the solution of
the equation
\begin{equation}
\lambda C_\textsf{SR}^{\textrm{(Ideal)}}=\lambda C_\textsf{SD}+(1-\lambda)J^\textrm{(Ideal)}_{\textrm{IA}}
\label{f1eqf2}
\end{equation}
with
\begin{equation}\label{lmd1}
\lambda_1=\frac{-\frac{\mathcal{W}(-\frac{a}{c}e^{-b-\frac{a}{c}})}{a}-\frac{1}{c}}{1-\frac{\mathcal{W}(-\frac{a}{c}e^{-b-\frac{a}{c}})}{a}-\frac{1}{c}},
\end{equation}
and
$\lambda_2$ is the $\lambda$-coordinate of the stagnation point of $f_1(\lambda)$ with
\begin{equation}\label{lmd2}
\lambda_2=\frac{e^{\mathcal{W}(\frac{c-1}{e})+1}-1}{e^{\mathcal{W}(\frac{c-1}{e})+1}+c-1},
\end{equation}
where $\mathcal{W}(x)$ is \emph{Lambert-W function}, which is defined by the equation
$x=\mathcal{W}(x)e^{\mathcal{W}(x)}$.
\end{thm}
\begin{IEEEproof}
The proof can be referred to Appendix A and Figure \ref{Fig2} gives the geometric interpretation.
\end{IEEEproof}

It is worth noting that since $\mathcal{W}(x)$ is multivalued (except at 0) and has two branches in real number field, i.e., $\mathcal{W}_0(x)$ and $\mathcal{W}_{-1}(x)$. Thus, we need to determine which branch should be adopted in (\ref{lmd1}) and (\ref{lmd2}). Specifically, for (\ref{lmd1}) and (\ref{lmd2}), $\mathcal{W}_{-1}(x)$ and $\mathcal{W}_0(x)$ should be adopted respectively. The reason and analysis can be found in Appendix C.

By using Theorem 1, the optimal time fraction $\lambda^*$ is obtained and further the maximum achievable rate can be calculated by using (\ref{RI}).


\subsubsection{{Ideal protocol with EA method}}
Now we consider the {ideal protocol with EA method}, for which we present the following Lemma 3 at first.
\begin{lem}
{Let $f_2(\lambda)=\lambda C_\textsf{SD}+(1-\lambda)J^\textrm{(Ideal)}_{\textrm{EA}}$. $f_2(\lambda)$ is a concave function w.r.t variable $\lambda$ for $\lambda \in (0,1)$.}
\label{L3}
\end{lem}
\begin{IEEEproof}
This lemma can be proved by the first-order derivative and the second-order derivative  of $f_2(\lambda)$, which is similar to {the proof of} Lemma 2. Therefore, the detailed proof is omitted here.
\end{IEEEproof}

Based on  Lemma 3, the following theorem is  given to calculate the optimal $\lambda^*$ of EA method for Problem \textbf{P1}.
\begin{thm}
\label{TIdealEA}
The optimal solution $\lambda^*$ of  EA method of Problem \textbf{P1} to achieve the maximum achievable rate can be calculated by
$\lambda^*=\max\{\lambda_1,\lambda_2\}$,
where $\lambda_1$ is the solution of
the equation
\begin{equation}
\lambda C_\textsf{SR}^{\textrm{(Ideal)}}=\lambda C_\textsf{SD}+(1-\lambda)J^\textrm{(Ideal)}_{\textrm{EA}}
\label{f1eqf3}
\end{equation}
with
\begin{equation}\label{optimallambdaEC1}
\lambda_1=\frac{-\frac{\mathcal{W}(-\frac{a}{c}e^{-\frac{a(1+m)}{c}})}{a}-\frac{1+m}{c}}{1-\frac{\mathcal{W}(-\frac{a}{c}e^{-\frac{a(1+m)}{c}})}{a}-\frac{1+m}{c}},
\end{equation}
and
$\lambda_2$ is the stagnation point of $f_2(\lambda)$ with
\begin{equation}\label{optimallambdaEC2}
\lambda_2=\frac{e^{\mathcal{W}(\frac{c-(1+m)}{e^{(1+b)}})+b+1}-(1+m)}{e^{\mathcal{W}(\frac{c-(1+m)}{e^{(1+b)}})+b+1}+c-(1+m)}.
\end{equation}
\end{thm}
\begin{IEEEproof}
The proof can be found in Appendix B.
\end{IEEEproof}

Similar to IA method, we also need to determine which branch should be adopted in (\ref{optimallambdaEC1}) and (\ref{optimallambdaEC2}). Specifically, for (\ref{optimallambdaEC1}) and (\ref{optimallambdaEC2}), $\mathcal{W}_{-1}(x)$ and $\mathcal{W}_0(x)$ should be adopted respectively. The reason and analysis can be found in Appendix C.
Therefore, for EA method, the optimal $\lambda^*$ is obtained by Theorem 2, and the maximum achievable rate can be {calculated} by using (\ref{RI}).

\subsection{PS Protocol}

To explore the system performance limit of PS protocol, we formulate the following optimization problem to jointly determine the optimal values of parameters  $\lambda$ and $\rho$ for both IA and EA as
\begin{flalign}
\textbf{P2}:\,\,\max\limits_{\lambda,\rho}\,\,&R^\textrm{(PS)}=\min\{\lambda C_\textsf{SR}^\textrm{(PS)}, \lambda C_\textsf{SD}+(1-\lambda)J^\textrm{(PS)}\}\\
\textrm{s.t.}\,\,&0<\lambda<1,\,\,0\leq\rho<\rho^{\textrm{th}},\nonumber
\end{flalign}
where $\rho^{\textrm{th}}=1-\frac{{H_\textsf{SD}\sigma^2_\textrm{b}}}{H_\textsf{SR}{\sigma^2_\textsf{D}}-{H_\textsf{SD}\sigma^2_\textrm{a}}}$.

{It can be observed that} Problem \textbf{P2} is non-convex, which is
difficult to solve directly.  However, by comparing Problem \textbf{P2}
with Problem \textbf{P1}, fortunately, we found that for a fixed
$\rho \in [0,\rho^{\textrm{th}})$, the Problem \textbf{P2} with the
variable $\lambda$ have the same structure with that of Problem
\textbf{P1}. This indicates that for a fixed $\rho \in
[0,\rho^{\textrm{th}})$, the optimal explicit solution $\lambda^*$
of Problem \textbf{P2} can be obtained with the similar {solving
method to} Problem \textbf{P1}. Based on this observation, we propose a
two-step algorithm to solve Problem \textbf{P2} as follows.

\textbf{Step 1:} For a given power splitting ratio $\rho$,
calculating the optimal time fraction $\lambda^*$. The detail operation of Step 1 is described as follows.  With a given
$\rho \in [0,\rho^{\textrm{th}})$, Problem \textbf{P2} is reduced to
be
\[
\begin{array}{l l}
\textbf{P2$'$}:&\max\limits_{\lambda}\,\,\min\{\lambda C_\textsf{SR}^\textrm{(PS)}, \lambda C_\textsf{SD}+(1-\lambda)J^\textrm{(PS)}\}\\
&\textrm{s.t.}\ \ 0<\lambda<1.
\end{array}
\]

Defining two new notations related to $\rho$, $a'\triangleq
C_\textsf{SR}^{\textrm{(PS)}}-C_\textsf{SD}$,  $c'\triangleq
\frac{\eta \rho
H_\textsf{SR}H_\textsf{RD}P_\textsf{S}}{\sigma^2_\textsf{D}}$,
then we obtain the following two corollaries associated with the
optimal $\lambda^*$ for PS protocol with IA and EA methods,
respectively.

\begin{cor}
With a given $\rho \in [0,\rho^{\textrm{th}})$, the optimal
solution $\lambda^*$ of Problem \textbf{P2$'$} for IA method is
$\lambda^*=\max\{\lambda_1,\lambda_2\}$,
where
$
\lambda_1=\frac{-\frac{\mathcal{W}(-\frac{a'}{c'}e^{-b-\frac{a'}{c'}})}{a'}-\frac{1}{c'}}{1-\frac{\mathcal{W}(-\frac{a'}{c'}e^{-b-\frac{a'}{c'}})}{a}-\frac{1}{c'}},
$
and
$
\lambda_2=\frac{e^{\mathcal{W}(\frac{c'-1}{e})+1}-1}{e^{\mathcal{W}(\frac{c'-1}{e})+1}+c'-1}.
$
\end{cor}

\begin{cor}
With a given $\rho \in [0,\rho^{\textrm{th}})$,  the optimal
solution $\lambda^*$  of Problem \textbf{P2$'$} for EA method is
$
\lambda^*=\max\{\lambda_1,\lambda_2\},
$
where
$\lambda_1=\frac{-\frac{\mathcal{W}(-\frac{a'}{c'}e^{-\frac{a'(1+m)}{c'}})}{a'}-\frac{1+m}{c'}}{1-\frac{\mathcal{W}(-\frac{a'}{c'}e^{-\frac{a'(1+m)}{c'}})}{a'}-\frac{1+m}{c'}},
$
and
$
\lambda_2=\frac{e^{\mathcal{W}(\frac{c'-(1+m)}{e^{(1+b)}})+b+1}-(1+m)}{e^{\mathcal{W}(\frac{c'-(1+m)}{e^{(1+b)}})+b+1}+c'-(1+m)}.
$
\end{cor}

Since  Problem \textbf{P2$'$} has the same form with that of Problem
\textbf{P1}, Corollary 1 and Corollary 2 can be directly derived
from Theorem 1 and Theorem 2, respectively. Thus, we omit the
detailed proof of Corollary 1 and Corollary 2 here.

\textbf{Step 2:} Find the optimal power splitting ratio $\rho^*$. The detail operation of Step 2 is as follows.
It is seen that the obtained optimal time fraction $\lambda^*$ in
Step 1 for both IA and EA can be regarded as a function of $\rho$.
Thus, by substituting $\lambda^*(\rho)$ into Problem \textbf{P2},
Problem \textbf{P2} then is transformed {into} the following two
expressions for IA and EA methods, respectively. For IA method, it
is
\begin{flalign}
\max\limits_{\rho}\,\,&\min\left\{\lambda^* (a'+b), \lambda^* b+(1-\lambda^*)\Big(b+\mathcal{C}\big(\frac{c'}{1/\lambda^*-1}\big)\Big)\right\}\\
\textrm{s.t.}\,\,&0\leq\rho<\rho^{\textrm{th}},\nonumber
\end{flalign}
and for EA method, it is
\begin{flalign}
\max\limits_{\rho}\,\,&\min\left\{\lambda^* (a'+b), \lambda^* b+(1-\lambda^*)\Big(\mathcal{C}\big(m+\frac{c'}{1/\lambda^*-1}\big)\Big)\right\}\\
\textrm{s.t.}\,\,&0\leq\rho<\rho^{\textrm{th}}.\nonumber
\end{flalign}

Nevertheless, both objective functions of above two
optimization problems are neither convex nor concave and it is hard to obtain the explicit expression of the
optimal $\rho^*$ for both of them. As our goal is to explore the
performance limit of the systems, therefore, the linear
search method is adopted to find the optimal $\rho^*$ over $\rho \in
[0,\rho^{\textrm{th}})$ with a search step-size $\epsilon$.

\subsection{TS Protocol}

Similary, the uniform optimization problem associated with TS protocol for both IA and EA can be given by
\begin{flalign}
\textbf{P3}:\,\,\max\limits_{\alpha_1,\alpha_2,\alpha_3}\,\,&R^\textrm{(TS)}=\min\{\alpha_1 C_\textsf{SR}^\textrm{(TS)}, (\alpha_1+\alpha_2) C_\textsf{SD}+\alpha_3J^\textrm{(TS)}\}\\
\textrm{s.t.}\,\,&\alpha_1+\alpha_2+\alpha_3=1,\label{a1a2a3}\\
&\alpha_1,\alpha_2,\alpha_3\geq0.
\end{flalign}

To get the optimal solution of Problem \textbf{P3}, we first transform it to be a convex problem. By introducing an auxiliary variable $r$, we equivalently transform Problem \textbf{P3} into the following Problem
\textbf{P3$'$}.
\begin{flalign}\label{RT1}
\textbf{P3$'$}:\,\,\min_{r, \alpha_1, \alpha_2, \alpha_3}\,\,&\ -r\\
\textrm{s.t.} \,\,&r-(\alpha_1+\alpha_2) C_\textsf{SD}-\alpha_3J^\textrm{(TS)}\leq0\label{Cst0}\\
&r-\alpha_1 C_\textsf{SR}^\textrm{(TS)}\leq0\label{Cst1}\\
&\alpha_1+\alpha_2+\alpha_3=1\label{Cst2}\\
&\alpha_1,\alpha_2,\alpha_3\geq0\label{Cst3}.
\end{flalign}
Based on this, we get the following theorem for Problem
\textbf{P3$'$}.
\begin{thm}
{Problem \textbf{P3$'$} is a convex problem.}
\end{thm}
\begin{IEEEproof}
It can be seen that, the objective function and the constraints
(\ref{Cst1}), (\ref{Cst2}) and (\ref{Cst3}) of problem
\textbf{P3$'$} are linear w.r.t. variables $\alpha_1$, $\alpha_2$,
$\alpha_3$ and $r$. If we can prove that the constraint
(\ref{Cst0}) is a convex set, Theorem 3 is proved. To discuss the
convexity of constraint (\ref{Cst0}), we first consider the IA
case, where $J^\textrm{(TS)}=J^\textrm{(TS)}_{\textrm{IA}}$ in
constraint (\ref{Cst0}). Substituting  (\ref{PRT}) into
(\ref{JIC}), we have that
$
J^\textrm{(TS)}_{\textrm{IA}}=b+\log\left(1+\frac{c\alpha_2}{\alpha_3}\right).
$
Via a simple transformation, the constraint (\ref{Cst0}) is
written to be
$r-(\alpha_1+\alpha_2+\alpha_3)
C_\textsf{SD}-\alpha_3\log\left(1+\frac{c\alpha_2}{\alpha_3}\right)\leq0$.
Let $g(\alpha_2)=\log(1+{c\alpha_2})$, which is a concave function
w.r.t $\alpha_2$. Thus, its perspective function
$h(\alpha_2,\alpha_3)\triangleq \alpha_3g(\alpha_2/\alpha_3)$ is
also a concave function \cite{Boyd}. As a result,
$-h(\alpha_2,\alpha_3)$ is a convex function, which means that the
left-hand side of constraint (\ref{Cst0}) is convex. {So}, for
the IA case, Problem \textbf{P3$'$} is a convex problem. For the EA
case, where $J^\textrm{(TS)}=J^\textrm{(TS)}_{\textrm{EA}}$ in
constraint (\ref{Cst0}). Substituting  (\ref{PRT}) into
(\ref{JEC}), we have that $
J^\textrm{(TS)}_{\textrm{EA}}=\log(1+m+\frac{c\alpha_2}{\alpha_3}).
$ With some simple manipulation, the constraint (\ref{Cst0}) can be
expressed as
$r-(\alpha_1+\alpha_2)
C_\textsf{SD}-\alpha_3\log\left(1+m+\frac{c\alpha_2}{\alpha_3}\right)\leq0$.
Similar to the IA {case}, it also can be proved that the left-hand
side of the constraint (\ref{Cst0}) is convex. So for the EA case,
 Problem \textbf{P3$'$} is a convex problem. Theorem 3
is thus proved.
\end{IEEEproof}

Although Problem \textbf{P3$'$} is a
convex problem, it is still hard to derive the explicit expressions
of its optimal solutions by adopting some existing methods including
Karush-Kuhn-Tucker (KKT) conditions, because the three variables are
closely coupled with each other in (\ref{Cst0}). In this case, by using
some software tools, such as \emph{CVX}\cite{Boyd}, one can obtain
the numerical solution of it.
However, considering that numerical solution cannot provide deep insights for the system, in order to better
understand TS protocol, we present an alternative solution method.
Before it, we provide some theoretical
properties associated with Problem \textbf{P3}.

\begin{lem} The optimal solution ($\alpha_1^*,\alpha_2^*,\alpha_3^*$) of Problem \textbf{P3} satisfies that
\begin{equation}\label{P41}
\alpha_1^* C_\textsf{SR}^\textrm{(TS)}=(\alpha_1^*+\alpha_2^*)
C_\textsf{SD}+\alpha_3^*J^{\textrm{(TS)}*},
\end{equation}
where $J^{\textrm{(TS)}*}$ is obtained by substituting $\alpha_2^*$ and $\alpha_3^*$ into $J^{\textrm{(TS)}}$.
\label{LTS}
\end{lem}
\begin{IEEEproof}
{Lemma 4 can be proved by the contradiction.}
Firstly, we consider the IA method case. From (\ref{PRT}) and (\ref{JIC}), and by using $\alpha_1+\alpha_2+\alpha_3=1$ to eliminate the variable $\alpha_2$, the right-hand side of (\ref{P41}) can be transformed {into}
\begin{equation}\label{RTSright}
R^\textrm{(TS)}_{\textrm{right}}=(1-\alpha_3) C_\textsf{SD}+\alpha_3\left(C_\textsf{SD}+\log\left(1+\tfrac{c(1-\alpha_1-\alpha_3)}{\alpha_3}\right)\right).
\end{equation}
Suppose that at {its} optimality, the left-hand side of (\ref{P41}) is larger than its right-hand side. If we fix $\alpha_3$ and reduce $\alpha_1$, $R^\textrm{(TS)}_{\textrm{right}}$ will be increased, leading to a higher system end-to-end achievable information rate, which is contradictory with the assumption that the achievable rate is optimal. On the other hand, suppose at {its} optimality, the left-hand side of (\ref{P41}) is less than its right-hand side. If we increase $\alpha_1$ with $\alpha_3$ fixed, $R^\textrm{(TS)}_{\textrm{right}}$ will be decreased, leading to a lower system end-to-end information achievable rate, which also contradicts with the assumption that the achievable rate is optimal. In summary, it is obtained that when the system achieves its maximum information rate, the two terms in (\ref{RT}) should be equal.
{Likewise, by using a similar way, } the same conclusion on EA method can also  be proved. Therefore, Lemma 4 is proved.
\end{IEEEproof}

With the equation of Lemma 4 and by substituting $\alpha_2$ with
$\alpha_2=1-\alpha_1+\alpha_3$, Problem \textbf{P3} can be
equivalently transformed {into} the following Problem \textbf{P3$''$},
\begin{align}\label{P4}
\textbf{P3$''$}:\,\,\min_{\alpha_1, \alpha_3} \,\,&\alpha_1 C_\textsf{SR}^\textrm{(TS)}\\
\textrm{s.t.}\,\,&\alpha_1 C_\textsf{SR}^\textrm{(TS)}=(1-\alpha_3) C_\textsf{SD}+\alpha_3J^\textrm{(TS)}\label{P5TS}\\
&\alpha_1+\alpha_3\leq1\label{P42}\\
&\alpha_1,\alpha_3\geq0\label{P43}.
\end{align}
The
solution of Problem \textbf{P3$''$} is derived for IA case and EA case, respectively, as follows.
\subsubsection{{{TS protocol with IA method}}}
Firstly, we consider the IA method. In this case, according to (\ref{JIC}) and (\ref{PRT}),
(\ref{P5TS}) can be transformed {into}
$
\alpha_1 C_\textsf{SR}^\textrm{(TS)}=(1-\alpha_3)
C_\textsf{SD}+\alpha_3\left(C_\textsf{SD}+\mathcal{C}\big(\tfrac{c(1-\alpha_1-\alpha_3)}{\alpha_3}\big)\right).
$
Further, we have that
$
\frac{\alpha_1 C_\textsf{SR}^\textrm{(TS)}-C_\textsf{SD}}{\alpha_3}=\mathcal{C}\left(\frac{c(1-\alpha_1-\alpha_3)}{\alpha_3}\right).
$
With the Lambert-W function $\mathcal{W}(x)$, one can obtain $\alpha_1$ as
\begin{equation}\label{alpha1}
\alpha_1=-\frac{\alpha_3
\mathcal{W}(\theta(\alpha_3))}{C_\textsf{SR}^\textrm{(TS)}}
+1-\alpha_3+\frac{\alpha_3}{c},
\end{equation}
where $\alpha_1$ can be considered as a function of $\alpha_3$ and
$\theta(\alpha_3)=\frac{C_\textsf{SR}^\textrm{(TS)}}{c}e^{-\frac{C_\textsf{SD}}{\alpha_3}+\frac{C_\textsf{SR}^\textrm{(TS)}}{c}(\frac{c(1-\alpha_3)}{\alpha_3}+1)}$.
$\mathcal{W}(x)$ should adopt the branch of $\mathcal{W}_0(x)$ and detailed
reason is similar to that in Appendix C.  Then, substituting (\ref{alpha1}), one can transform Problem
\textbf{P3$''$} {into} Problem \textbf{P3$^\Delta$}
\begin{flalign}\label{P5}
\textbf{P3$^\Delta$}:\,\,\min_{\alpha_3}\,\,&\left(-\frac{\alpha_3 \mathcal{W}\left(\theta(\alpha_3)\right)}{C_\textsf{SR}^\textrm{(TS)}}+1-\alpha_3+\frac{\alpha_3}{c}\right)C_\textsf{SR}^\textrm{(TS)}\nonumber\\
\textrm{s.t.}\,\,&0<\alpha_3\leq1\nonumber.
\end{flalign}

In Problem \textbf{P3$^\Delta$}, there is only one variable. As its objective function is neither convex nor concave, we adopt a linear search method  over $\alpha_3 \in (0,1]$
with a search step-size $\epsilon$ to find the optimal $\alpha_3^*$. Once $\alpha_3^*$ is obtained, $\alpha_1^*$ and
$\alpha_2$ can be calculated by using (\ref{alpha1}) and
(\ref{a1a2a3}).

\subsubsection{{TS protocol with EA method}}
For the EA method case, according to (\ref{JEC}) and (\ref{P5TS}),
we have that
\begin{equation}\label{TSalpha}
\alpha_1=-\frac{\alpha_3
\mathcal{W}(\phi(\alpha_3))}{C_\textsf{SR}^\textrm{(TS)}}+1-\alpha_3+\frac{(1+m)\alpha_3}{c},
\end{equation}
where
$\phi(\alpha_3)=\frac{C_\textsf{SR}^\textrm{(TS)}}{c}e^{-\frac{(1-\alpha_3)C_\textsf{SD}}{\alpha_3}+\frac{C_\textsf{SR}^\textrm{(TS)}}{c}(\frac{c(1-\alpha_3)}{\alpha_3}+1+m)}$.
So Problem \textbf{P3$''$} can be transformed {into} the following
problem (\textbf{P3$^{\sharp}$})
\begin{flalign}\label{P6}
\min_{\alpha_3}\,\,&\left(-\frac{\alpha_3 \mathcal{W}(\phi(\alpha_3))}{C_\textsf{SR}^\textrm{(TS)}}+1-\alpha_3+\frac{(1+m)\alpha_3}{c}\right)C_\textsf{SR}^\textrm{(TS)} \\
\textrm{s.t.}\,\,&0<\alpha_3\leq1\label{P61}.
\end{flalign}

Similar to Problem \textbf{P3$^\Delta$}, there is also only one variable
in Problem \textbf{P3$^{\sharp}$} and the objective function of
Problem \textbf{P3$^{\sharp}$} is also neither convex nor concave.
The linear search method over $\alpha_3 \in (0,1]$ with
a step-size $\epsilon$ also can be used to find the optimal $\alpha_3^*$. Once $\alpha_3^*$ is obtained, $\alpha_1^*$ and
$\alpha_2^*$ can be calculated {by following} (\ref{TSalpha}) and
(\ref{a1a2a3}). Note that, although the explicit solution of Problem \textbf{P3} still cannot be obtained with the alterative method, with it we get the explicit functional relationship of $\alpha_1$ and $\alpha_3$, which may help to better understand the TS protocol.

\subsection{Complexity Analysis}
In this subsection, we analyze the computational complexity of the proposed solution
methods of the three protocols. For the ideal protocol, due to the
obtained explicit expression of the optimal achievable rate, the
computational complexity is $O(1)$. For the PS and TS protocols, as the linear search method is adopted, the
computational complexity is $O(1/\epsilon)$.

\section{Results and Discussions}
\begin{figure}[t]
\centering
\includegraphics[width=0.40\textwidth]{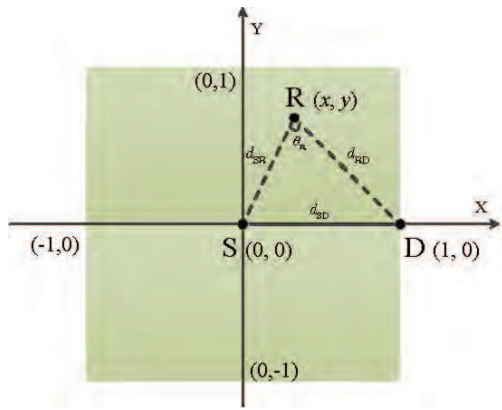}
\caption{The exemplary topologies for three-node relay networks.}
\label{Cordinate}
\end{figure}
This section provides some numerical results to discuss the system performance of our proposed three relaying protocols. In the simulations, we consider a typical three node network model, as shown in Figure  \ref{Cordinate}, where \textsf{S} is placed at the origin with the coordinate $(0,0)$ on the coordinate plane. \textsf{D} is place on the positive axis of $x$-axis. The distance between \textsf{S} and \textsf{D} is normalized to be 1, so the coordinate of \textsf{D} $(1,0)$. \textsf{R} is able to be positioned at any point {with coordinate $(x, y)$ of the square area satisfying that} $x\in [-1,1]$ and $y\in [-1,1]$, i.e., {the points within} the green area in Figure \ref{Cordinate}. All the channels are modeled to be flat Rayleigh fading channels and the noise power are normalized to be $\sigma_\textrm{a}^2=\sigma_\textrm{b}^2=1$. Thus,
$\sigma_\textsf{R}^2=\sigma_\textsf{D}^2=2$. The channel gain $H_\textsf{SD}$ of \textsf{S-D} link is regarded as the reference {and} the channel gains of the \textsf{S-R} link and \textsf{R-D} link were generated by $H_\textsf{SR}=G_\textsf{SR} H_\textsf{SD}$ and $H_\textsf{RD}=G_\textsf{RD} H_\textsf{SD}$, where $G_\textsf{SR}$ and $G_\textsf{RD}$ denote the channel gain ratios of \textsf{S-R} link and \textsf{R-D} link w.r.t \textsf{S-D} link, respectively. The pathloss effect of the channel is considered, i.e., $H_{uv}=(d_{uv})^{-\kappa}$, where $d_{{uv}}$ is the distance between $u$ and $v$, and $\kappa$ is path-loss factor which is set to 4 in our simulations. According to the pathloss model and the geometrical relationship among the nodes and {with the model in} \cite{Ravanshid}, we have that
\begin{equation}\label{G}
\left\{
\begin{aligned}
G_\textsf{SR}&=[1+\zeta_\textsf{R}^2-2\zeta_\textsf{R}\cos(\theta_\textsf{R})]^{\kappa/2}, \\
G_\textsf{RD}&=G_\textsf{SR}/\zeta_\textsf{R}^\kappa,
\end{aligned} \right.
\end{equation}
where
$\zeta_\textsf{R}\doteq\frac{d_\textsf{RD}}{d_\textsf{SR}}$ and $\theta_\textsf{R}$ is the angle of lines \textsf{R-D} and \textsf{R-S}.
In addition, the EH efficiency $\eta$ is set to be 1 in order to explore the potential performance of PS and TS protocol, and the
updating step-size $\epsilon$ is set to 0.001.

\subsection{Performance Comparisons}
\begin{figure}
\begin{minipage}[t]{0.49\textwidth}
\centering
  \includegraphics[width=2.8in]{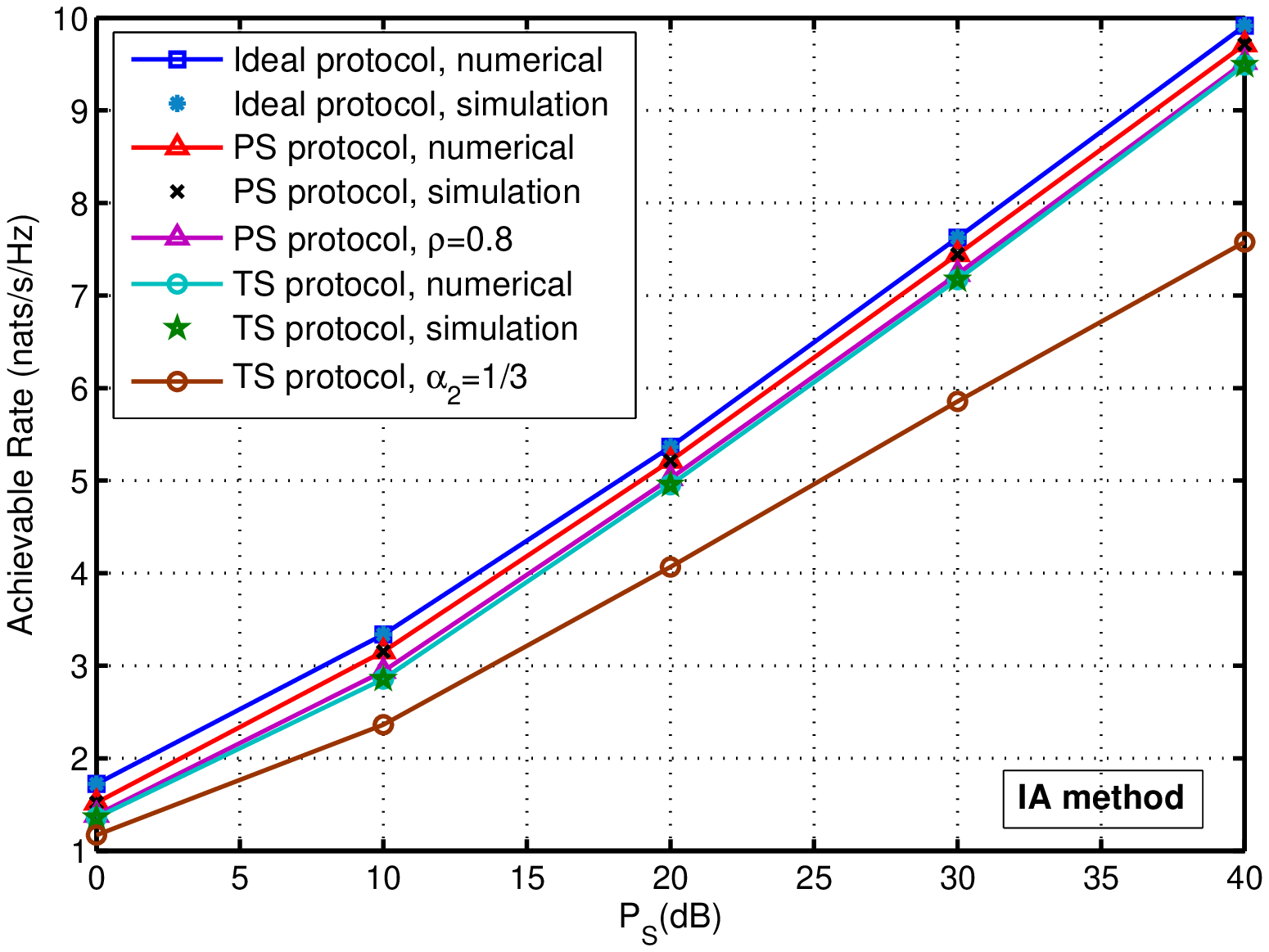}\\
\caption{Achievable rates v.s. $P_\textsf{S}$ for IA method.}
\label{Fig00}
\end{minipage}%
\begin{minipage}[t]{0.49\textwidth}
 \centering
\includegraphics[width=2.8in]{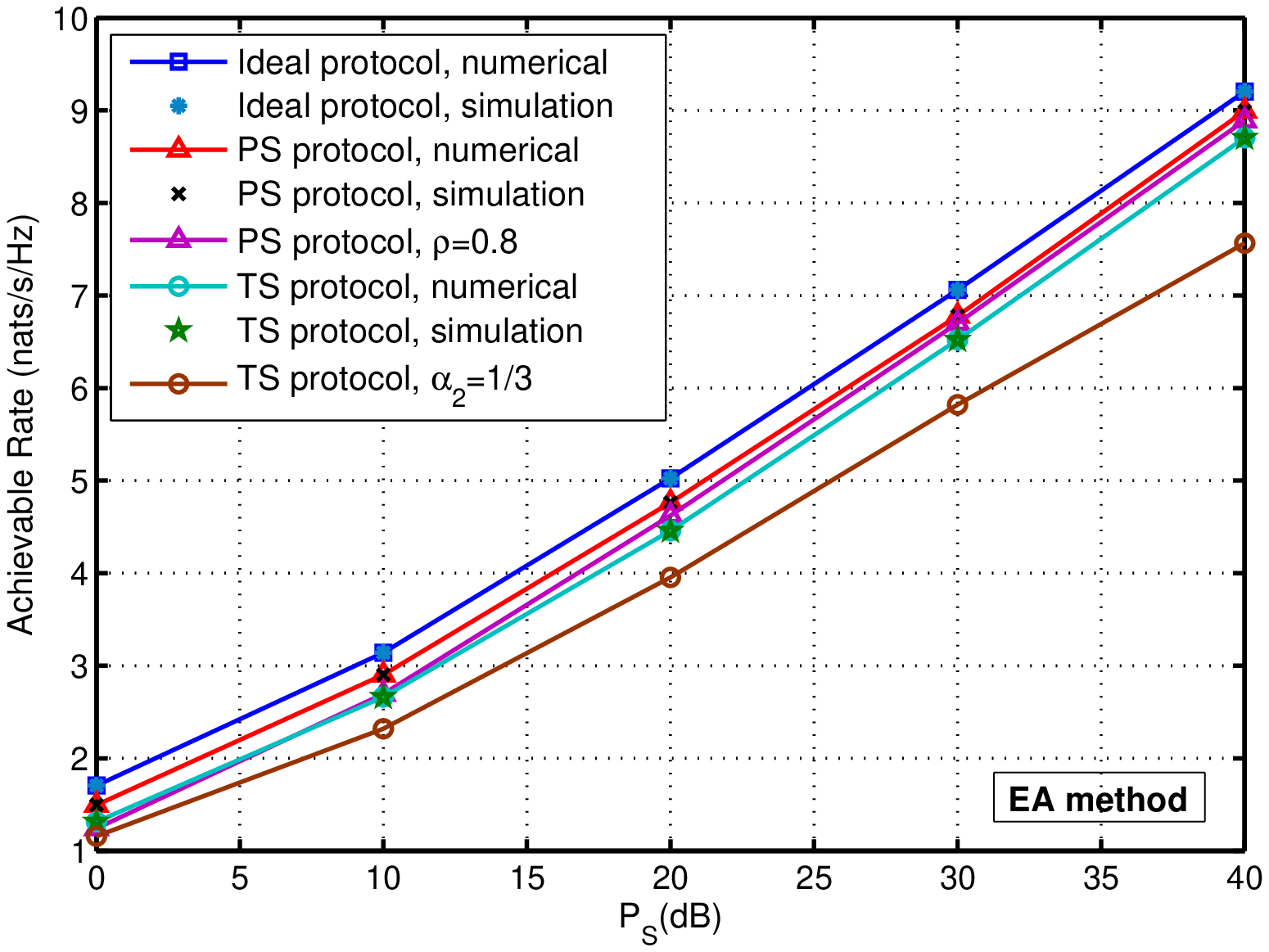}
\caption{Achievable rates v.s. $P_\textsf{S}$ for EA method.}
\label{Fig01}
\end{minipage}
\end{figure}

Figure  \ref{Fig00} and Figure  \ref{Fig01} present the achievable information rates of the three protocols versus the transmit power $P_\textsf{S}$ for IA and EA methods, respectively. In the simulations associated with the two figures, {$\theta_\textsf{R}$ is set to be $\pi$ and $\zeta_\textsf{R}$ is set to be $4/3$,} following which $G_\textsf{SR}=15$dB and $G_\textsf{RD}=10$dB.
In Figure  \ref{Fig00} and Figure  \ref{Fig01}, both the simulation results obtained by exhaustive computer search and the numerical results obtained by our analysis are plotted. It can be seen that the numerical ones match the simulation ones very well, which validates the correctness of our theoretical analysis and algorithms and also indicates that with our analytical results and designed algorithms, the optimum system performance can be achieved.
Moreover, the achievable information rates of two simply configured systems are also plotted as benchmark methods for comparison. For PS protocol, we consider the simple system configuration with $\rho=0.8$ and for TS protocol, we consider the simple system configuration with $\alpha_2=\frac{1}{3}$. It can be observed that via optimizing the system parameters such as $\rho$, $\lambda$, $\alpha_1$, $\alpha_2$ and $\alpha_3$, the system performance of PS and TS protocols can be greatly improved.
Besides, in both figures, it can be seen that PS protocol always outperforms TS protocol and the ideal protocol achieves the highest achievable information rate among the three protocols, which indicates that in practical RCed cooperative relaying system, PS protocol should be adopted in order to achieve the better system performance.

\begin{figure}
\begin{minipage}[t]{0.49\textwidth}
\centering
  \includegraphics[width=2.8in]{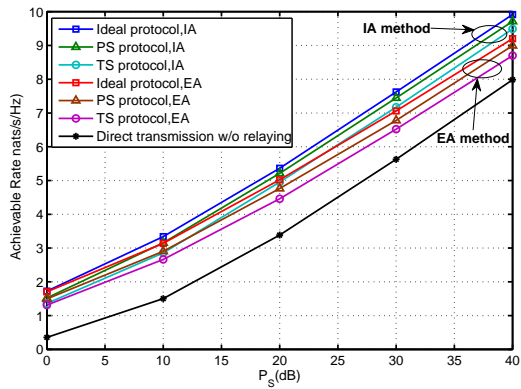}\\
\caption{Achievable rate comparison of IA and EA methods.}
\label{Fig3}
\end{minipage}%
\begin{minipage}[t]{0.49\textwidth}
 \centering
\includegraphics[width=2.8in]{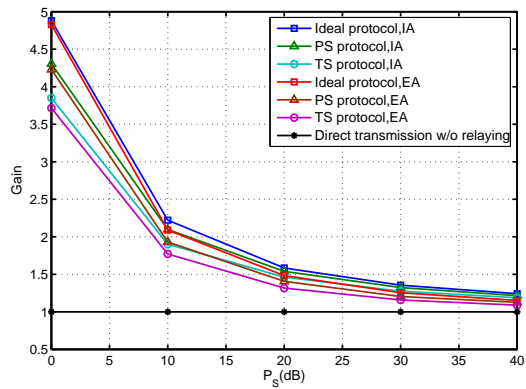}
\caption{Achievable rate gain to the direct-link transmission.}
\label{Figgain}
\end{minipage}
\end{figure}

To further compare the systems {with} different information receiving methods, we plot the achievable information rates of the three protocols with IA and EA methods versus the transmit power $P_\textsf{S}$ in Figure  \ref{Fig3}, where the  achievable information rate of the direct transmission over \textsf{S}-\textsf{D} link without relaying is {also} provided. It can be observed that for the same information receiving method the achievable information rate of the ideal protocol is the highest, and PS protocol is superior to TS protocol, which is consistent with the results in Figure \ref{Fig00} and Figure  \ref{Fig01}. It is also seen that for the same relaying protocol, the system with IA method always outperforms that with  EA method. Moreover, all relaying protocols achieve higher achievable information rates than the non-relaying direct link transmission, which means that by employing a SWIPT-enabled relay node with RC, the information transmission performance from a source to its destination can be greatly enhanced. In order to get more insights of the comparisons, we define the
performance gain as
$$
G\textrm{ain}^{\textrm{(Protocol)}}=\frac{R^{\textrm{(Protocol)}}}{C_\textsf{SD}},
$$
where $\textrm{Protocol}\in\{\textrm{Ideal,PS,TS}\}$ represents the ideal/PS/TS relaying protocols.
As $C_\textsf{SD}$  is the achievable information rate of the non-relaying direct link transmission, $G\textrm{ain}^{\textrm{(Protocol)}}$ actually describes the system performance enhancement of various relaying protocols compared to the non-relaying direct link transmission. Figure  \ref{Figgain} plots the achievable rate gains versus $P_\textsf{S}$. One can see that with the increment of $P_\textsf{S}$, the gains of all three relaying protocols gradually decrease, which approach 1 when $P_\textsf{S}\rightarrow \infty$. This observation indicates that, in relatively low SNR regime, SWIPT-enabled rateless code relaying is deserved to be employed to achieve much better system performance enhancement, while in relatively high SNR regime, employing the SWIPT-enabled rateless code relaying can only bring marginal performance enhancement.

\begin{figure}
\begin{minipage}[t]{0.49\textwidth}
\centering
  \includegraphics[width=2.8in]{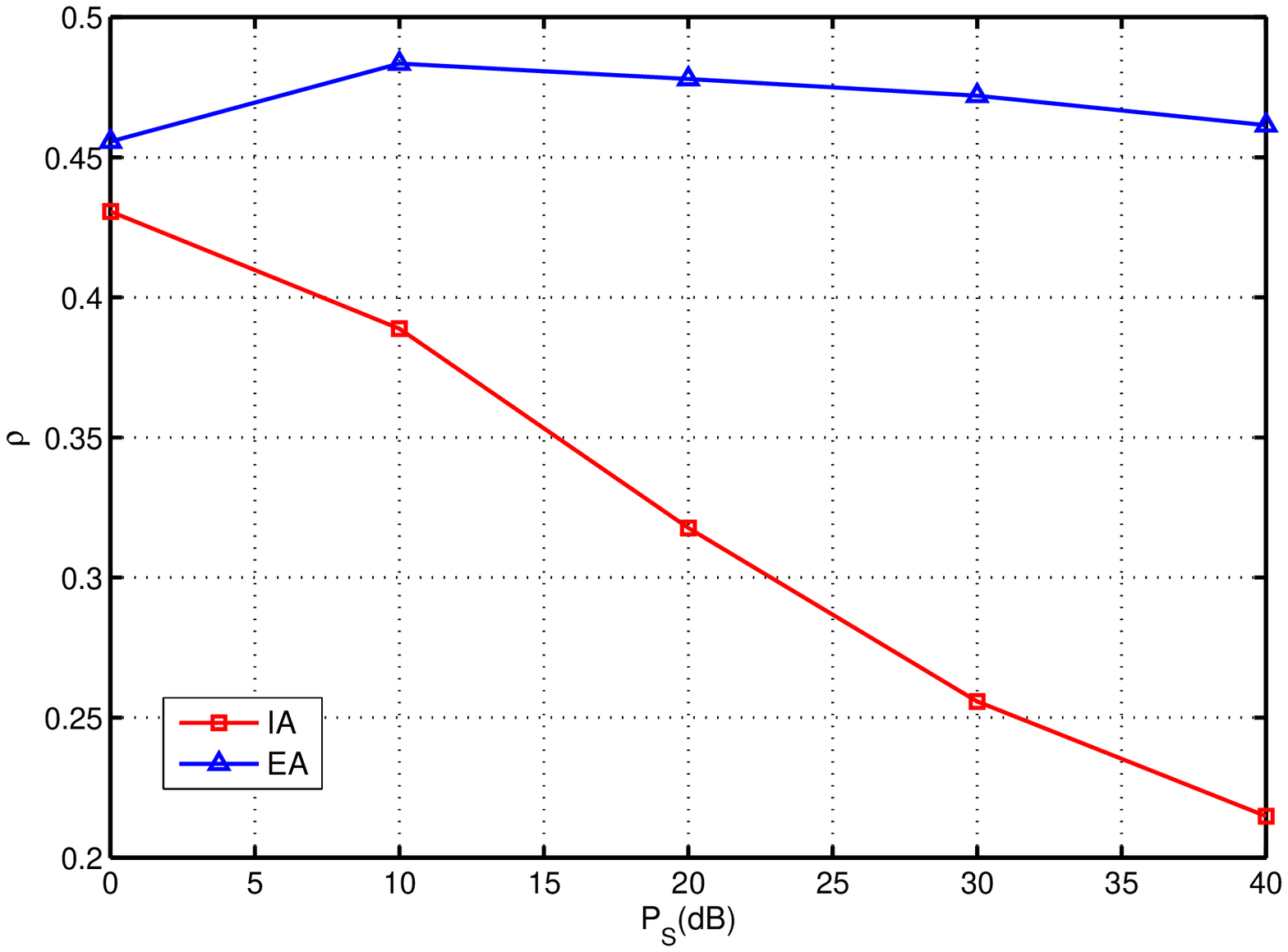}\\
\caption{The optimal $\rho^*$ of PS protocol vs $P_\textsf{S}$.}
\label{Fig4}
\end{minipage}%
\begin{minipage}[t]{0.49\textwidth}
 \centering
\includegraphics[width=2.8in]{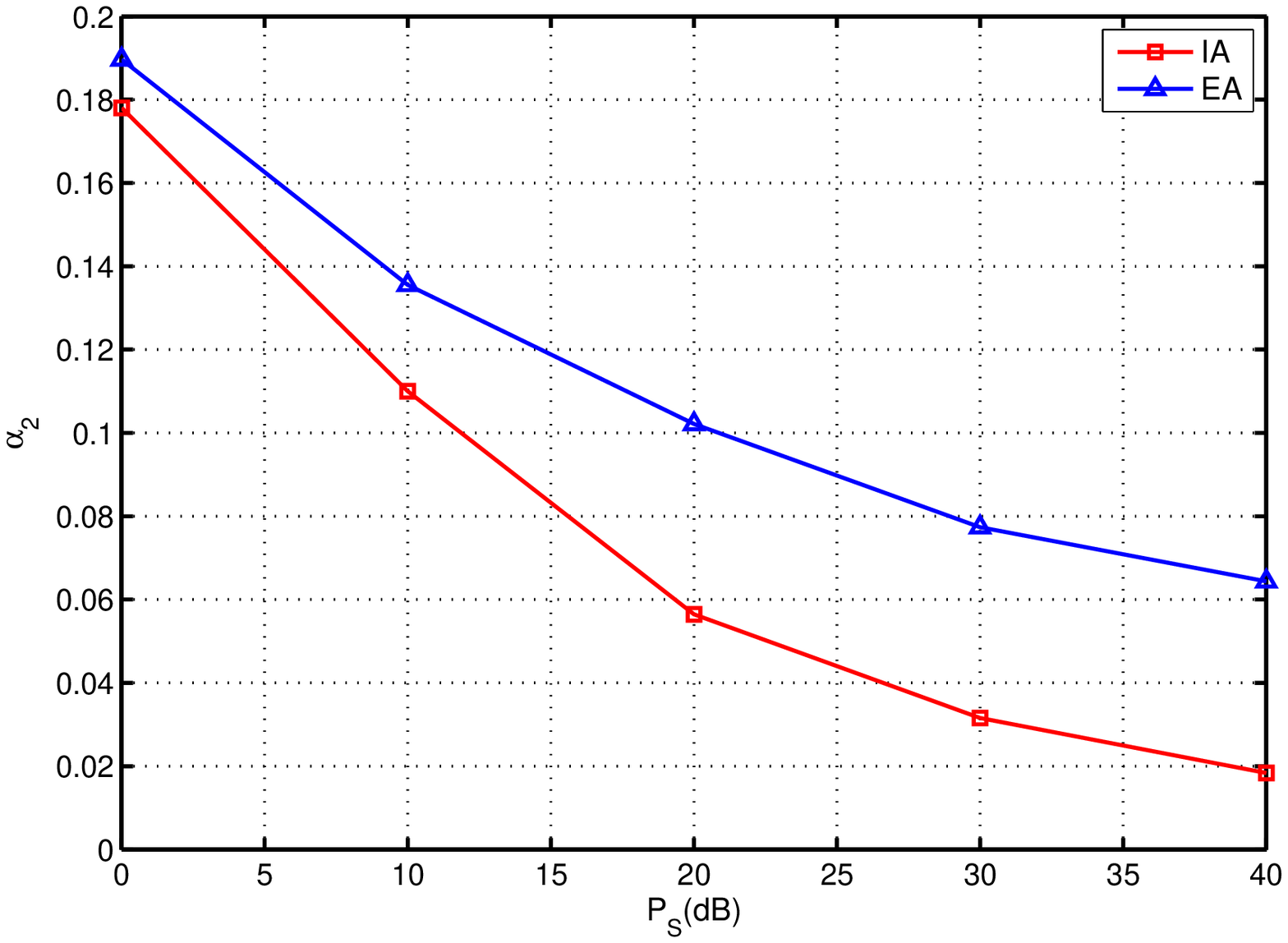}
\caption{The optimal $\alpha_2^*$ of TS protocol vs $P_\textsf{S}$..}
\label{Fig5}
\end{minipage}
\end{figure}

In Figure  \ref{Fig4} and Figure  \ref{Fig5}, we provide the optimal PS ratio $\rho$ of PS protocol and the optimal EH time fraction $\alpha_2$ of TS protocol versus $P_\textsf{S}$, respectively.  Figure  \ref{Fig4} shows that the PS ratio of EA method is higher than IA method and Figure  \ref{Fig5} shows  that the EH time fraction of IA method is lower than EA method. This means that, {compared with the protocol with IA method, in the protocol with EA method}, more received signals and more time should be assigned to perform EH for PS protocol and TS protocol, respectively.

\subsection{The Effects of direct link and RC on system performance}
In this subsection, we provide some numerical results to discuss the effects of direct link and RC on system performance. The network topology adopted {in the simulations} is shown in Figure  \ref{Cordinate}. Since IA information receiving method is always superior to EA method as {stated by} Figure \ref{Fig3} and Figure \ref{Figgain}, in the following simulations, we only take the IA method as a representative.

\begin{figure}
\begin{minipage}[t]{0.49\textwidth}
\centering
  \includegraphics[width=2.8in]{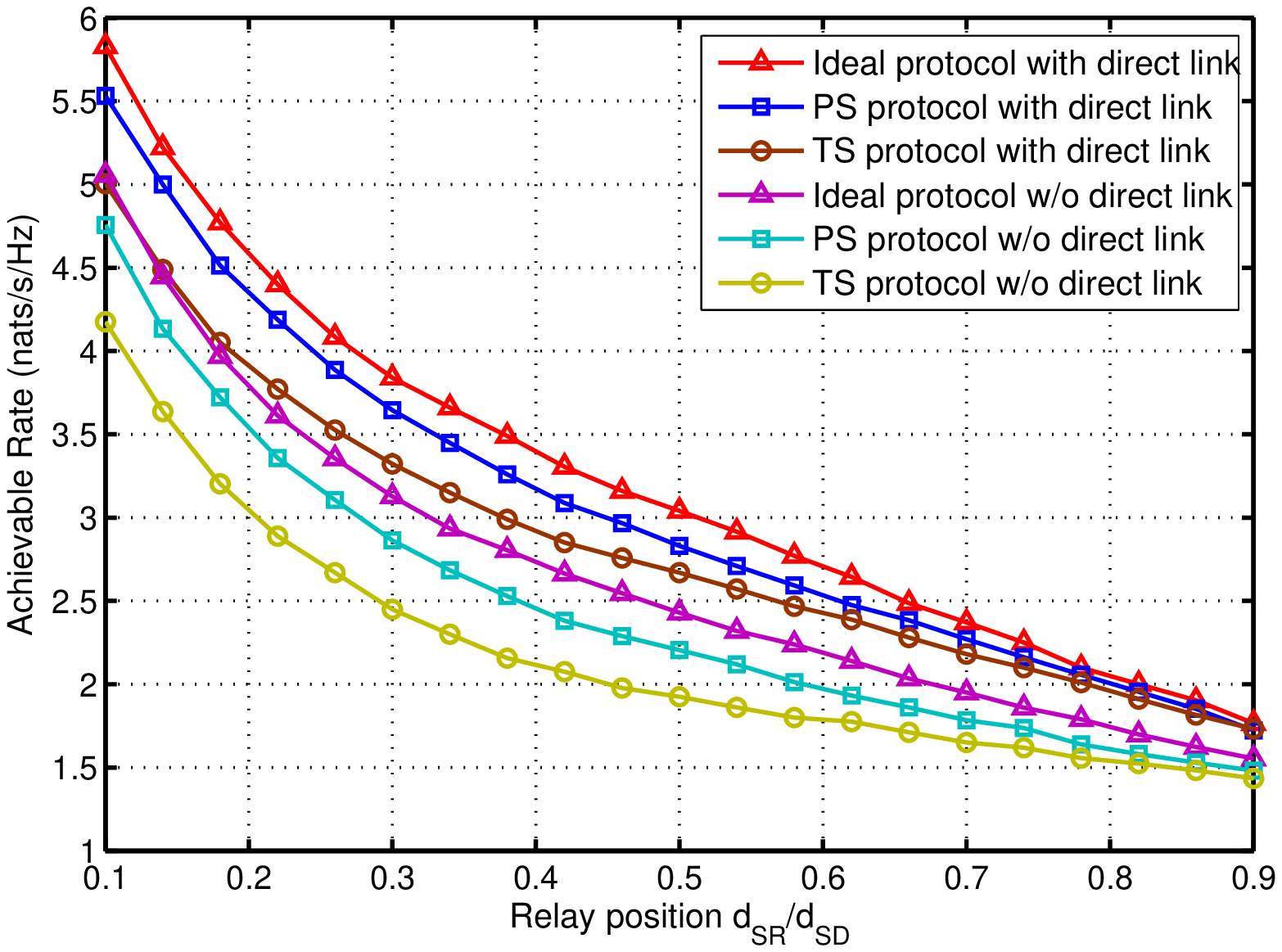}\\
\caption{The effects of direct link.}
\label{Figdirvsnodir}
\end{minipage}%
\begin{minipage}[t]{0.49\textwidth}
 \centering
\includegraphics[width=2.8in]{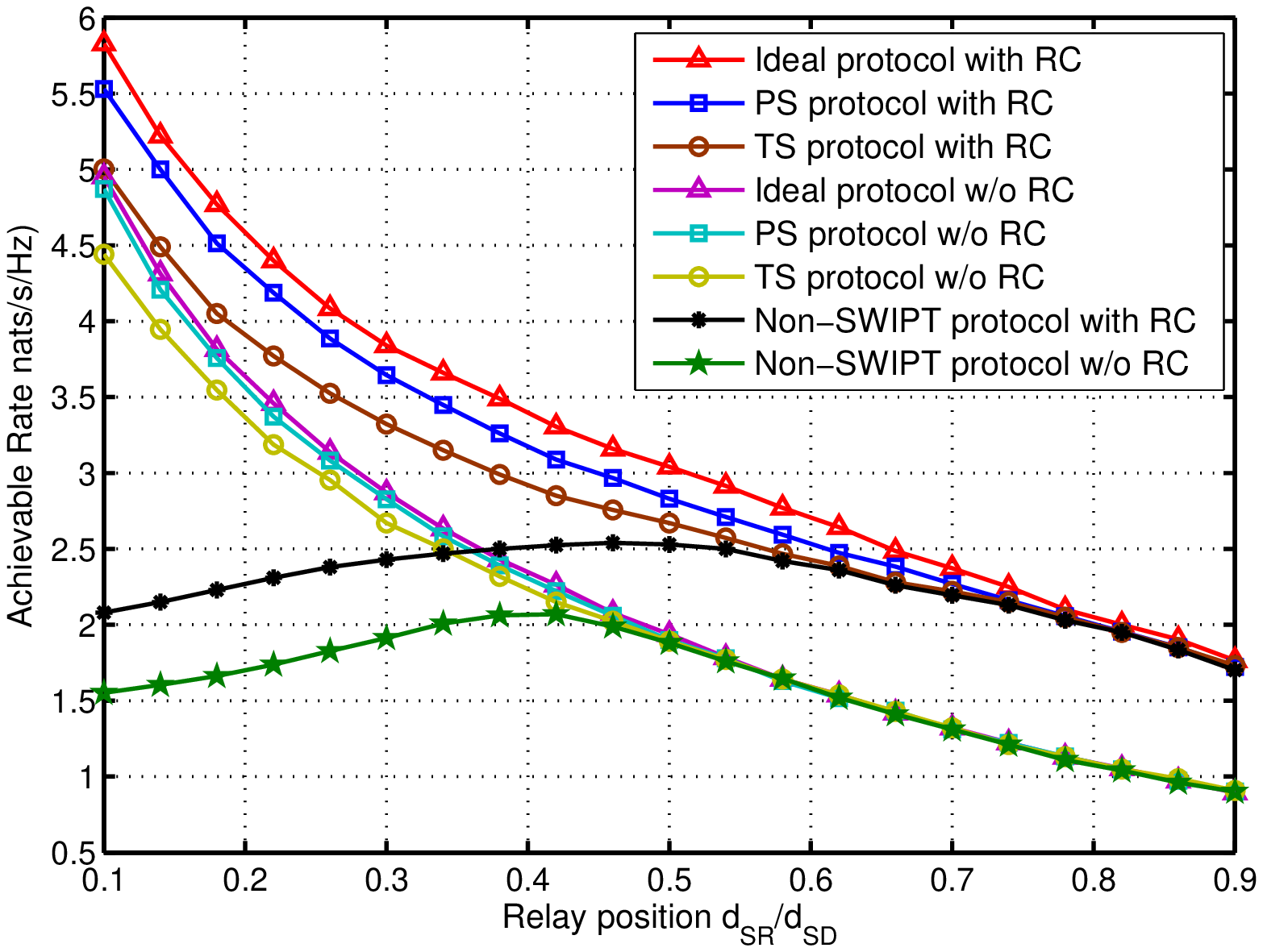}
\caption{{The effects of Rateless code.}}
\label{Figconv}
\end{minipage}
\end{figure}

In Figure  \ref{Figdirvsnodir}, we compare the performances of the ideal protocol, the PS protocol and the TS protocol in the system with \textsf{S}-\textsf{D} direct {link to those with} no direct link.  {$P_\textsf{S}$ and $\theta_\textsf{R}$ are set to be 10dB and $\pi$, respectively. The} relay is moved from \textsf{S} towards \textsf{D} on the line between \textsf{S} and \textsf{D}. $d_\textsf{SR}/d_\textsf{SD}$ is used to describe the relay position. It can be seen that all three protocols achieve much better performance in the system with \textsf{S}-\textsf{D} direct link than in the system with no direct link. This means that in the SWIPT-enabled RCed relaying system, with the \textsf{S}-\textsf{D} direct link cooperative transmission, the system performance can be greatly enhanced.

In Figure  \ref{Figconv}, we compare the ideal protocol, the PS protocol and the TS protocol with RC with their corresponding protocols without RC, where the performances of the conventional non-SWIPT protocols with RC and without RC are also plotted.
For fair comparison, the total available power of the conventional non-SWIPT protocol is assumed to be same with that of the three SWIPT-enabled protocols.
{Specifically}, for SWIPT-enabled protocols, the supplied total power of the system is $P_\textsf{S}$. For the non-SWIPT protocol with RC, the power at \textsf{S} an \textsf{R} are denoted as $\widetilde{P}_\textsf{S}, \widetilde{P}_\textsf{R}$, {which are set {to satisfy} that} $\widetilde{P}_\textsf{S}+(1-\lambda)\widetilde{P}_\textsf{R} = P_\textsf{S}$, and the maximum achievable information rate of the non-SWIPT protocol with RC is obtained by solving the following optimization problem.
\begin{flalign}\label{XX}
\max\limits_{\lambda}\,\,&\min\left\{\lambda C_\textsf{SR}^{\textrm{(NSRC)}}, \lambda C_\textsf{SD}^\textrm{(NSRC)}+(1-\lambda)J^\textrm{(NSRC)}\right\}\\
\textrm{s.t. }\,\,&0<\lambda<1,\nonumber\\
&\widetilde{P}_\textsf{S}+(1-\lambda)\widetilde{P}_\textsf{R}\leq P_\textsf{S},\nonumber
\end{flalign}
where $
C_\textsf{SR}^{\textrm{(NSRC)}}=\mathcal{C}\left(\frac{\widetilde{P}_\textsf{S}H_\textsf{SR}}{\sigma^2_\textsf{R}}\right),
$
$
C_\textsf{SD}^\textrm{(NSRC)}=\mathcal{C}\left(\frac{\widetilde{P}_\textsf{S}H_\textsf{SD}}{\sigma^2_\textsf{D}}\right)
$.
When IA method is adopted,
$
J^\textrm{(NSRC)}=J^\textrm{(NSRC)}_{\textrm{IA}}=\mathcal{C}\left(\frac{\widetilde{P}_\textsf{S} H_\textsf{SD}}{\sigma^2_\textsf{D}}\right)+\mathcal{C}\left(\frac{\widetilde{P}_\textsf{R}H_\textsf{RD}}{\sigma^2_\textsf{D}}\right),
$
and when EA method is adopted,
$
J^\textrm{(NSRC)}=J^\textrm{(NSRC)}_{\textrm{EA}}=\mathcal{C}\Big(\frac{\widetilde{P}_\textsf{S} H_\textsf{SD}}{\sigma^2_\textsf{D}}+$
$\frac{\widetilde{P}_\textsf{R}H_\textsf{RD}}{\sigma^2_\textsf{D}}\Big).
$
To solve this problem, we firstly fix $\lambda$ and solve the corresponding convex problem {w.r.t. $\widetilde{P}_\textsf{S}$ and $\widetilde{P}_\textsf{R}$}, and then get its optimal solution {$\lambda^*$ by linear} search. For the non-SWIPT protocol without RC, the durations of the two transmission phases are assumed to be equal and the signals from \textsf{S} and \textsf{D} are decoded by using MRC method. To explore its maximum achievable rate, an optimization problem can be formulated by letting $\lambda$ be 1/2 and substituting  $J^\textrm{(NSRC)}$ with $J^\textrm{(NSRC)}_{\textrm{EA}}$ into (\ref{XX}). Since the problem is convex, we solve it by CVX tools. {The comparison results are shown in Figure \ref{Figconv}. }

From Figure  \ref{Figconv}, it can be seen that in both the RCed and non-RCed relaying systems, SWIPT-enabled protocols always outperforms traditional non-SWIPT protocols. Moreover, when \textsf{R} is placed close to \textsf{D}, they achieve {a very} similar performance, because in this case, the energy harvested by \textsf{R} becomes much less than that in the case when \textsf{R} is placed close to \textsf{S}. Besides, it also shows that for the SWIPT-enabled relaying protocols, when \textsf{R} is placed close to \textsf{S}, the best system performance is achieved  while in traditional non-SWIPT relaying protocols, when \textsf{R} is placed at the source side {close to} the middle point on the \textsf{S-D} link, the best system performance can be achieved.



\subsection{The Effects of Relay Position on System Performance}
\begin{figure}
\begin{minipage}[t]{0.49\textwidth}
\centering
  \includegraphics[width=2.8in]{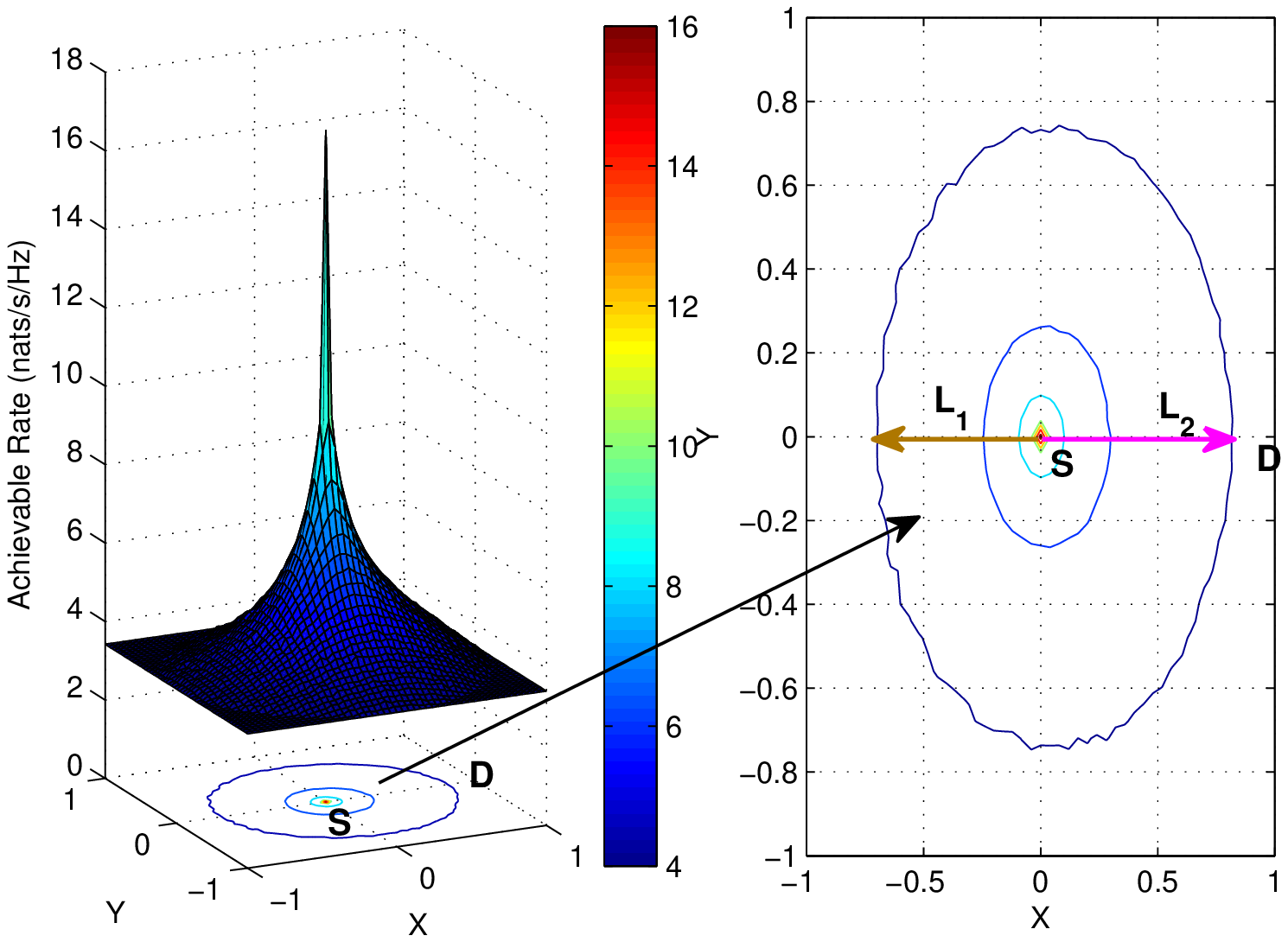}\\
\caption{The effect of relay position on Ideal protocol.}
\label{FigIdealIA}
\end{minipage}%
\begin{minipage}[t]{0.49\textwidth}
 \centering
\includegraphics[width=2.8in]{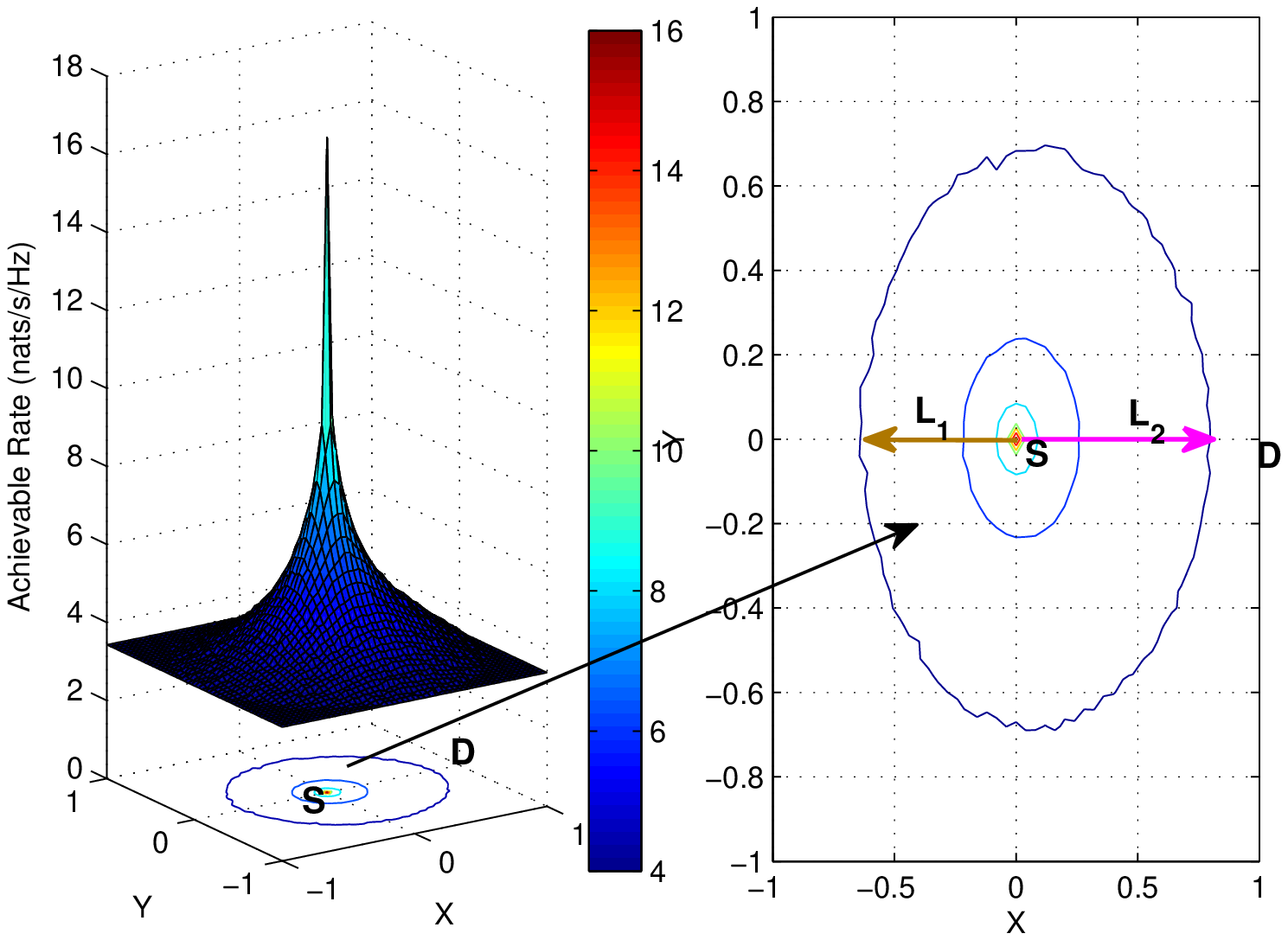}
\caption{The effect of relay position on PS protocol}
\label{FigPSIA}
\end{minipage}
\end{figure}

\begin{figure}[t]
\centering
\includegraphics[width=0.45\textwidth]{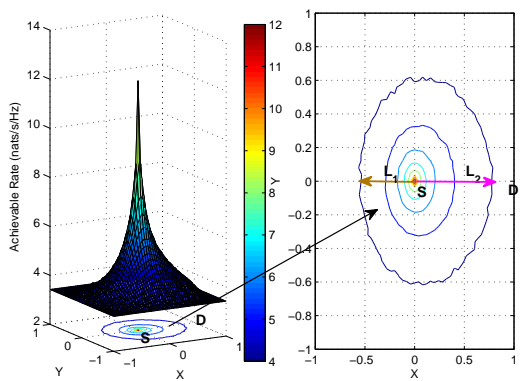}
\caption{{The effect of relay position on TS protocol}}
\label{FigTSIA}
\end{figure}

In Figure  \ref{FigIdealIA}, Figure  \ref{FigPSIA} and Figure  \ref{FigTSIA}, we plot the 3-dimension results and its contours of the achievable information rate of the ideal, the PS and the TS protocols, respectively. $P_\textsf{S}$ is set to 20dB. From the 3D results in the three figures, it can be seen that the closer the relay is located to the source, the better system performance can be achieved. Moreover, from the contours of these figures, {one can observe that when the relay is placed to the left side of the source, performance gain also can be achieved}. However, in the figures, it is shown that $L_1<L_2$, which means that for a given distance between the source and the relay, {positioning the relay at the right side of the source is better than positioning it at the left side of the source}.

\section{Conclusion}
This paper investigated the SWIPT in cooperative relay network{s with rateless codes}. Three relaying protocols were proposed for such a SWIPT-enabled relay network. To explore the system performance, three optimization problems were formulated to maximize the achievable information rate of the relaying protocols. Some analytical results and efficient algorithms were designed to solve the problems. Simulation results showed that with the same SWIPT receiver, the achievable information rate of IA system is superior to EA system, while for the same information receiving  strategy, the achievable information rate of PS protocol outperforms TS protocol. It also showed that compared with convectional non-SWIPT and non-RCed system, by employing SWIPT and RC, the system performance can be greatly improved. Besides, we also discussed the effects of the source-destination direct link and the relay position on system performance and provided some useful insights.

\appendices
\section{{Proof of Theorem 1}}
\begin{IEEEproof}
Firsly, we prove the equation $\lambda C_\textsf{SR}^{\textrm{(Ideal)}}=\lambda C_\textsf{SD}+(1-\lambda)J^\textrm{(Ideal)}_{\textrm{IA}}$ has one and only one solution in the interval (0,1). For this, it only need to prove that in (0,1), the corresponding curves of two functions $f_3(\lambda)=\lambda C_\textsf{SR}^{\textrm{(Ideal)}}$ and $f_1(\lambda)=\lambda C_\textsf{SD}+(1-\lambda)J^\textrm{(Ideal)}_{\textrm{IA}}$ have one and only one intersection point. I can be derived that
$\lim_{\lambda\rightarrow0}f_1(\lambda)=\lim_{\lambda\rightarrow1}f_1(\lambda)=C_\textsf{SD}$.
By introducing an auxiliary function, we have
\begin{equation}\label{F1}
F_1(\lambda)=\left \{
\begin{array}{ll}
C_\textsf{SD},& \lambda=0 \ \textrm{or} \ \lambda=1\\
f_1(\lambda),&0<\lambda<1\\
\end{array}\right.
\end{equation}
Let
$
F(\lambda)=f_3(\lambda)-F_1(\lambda).
$
It is obtained that $F(0)=-C_\textsf{SD}<0$ and $F(1)=C^\textrm{(Ideal)}_\textsf{SR}-C_\textsf{SD}>0$. Then using zero-point theorem, in the open interval (0,1), there exists a zero point for the function $F(\lambda)$, and thus (\ref{f1eqf2}) certainly has one solution within (0,1). Denote the solution of (\ref{f1eqf2}) as $\lambda_1$.
Now, we prove that (\ref{f1eqf2}) has only one solution in (0,1). $F_1(\lambda)$ is a strictly concave function. By using the property of concave function and combining $F_1(0)=F_1(1)$, we have that $F_1(\lambda)$ is first {increasing and then decreasing} function in (0,1). Let the $\lambda$-coordinate of the stagnation point of $f_1(\lambda)$ be $\lambda_2$. Consider two cases, that is, the case that the $\lambda$-coordinate of  the intersection point $\lambda_1$ is located at $(\lambda_2,1)$ (i.e., Case 1, $\lambda_1\geq\lambda_2$) and the case $(0,\lambda_2)$ (i.e., Case 2, $\lambda_1<\lambda_2$), as illustrated in Figure  \ref{Fig2}. Using the property of concave function, we prove that (\ref{f1eqf2}) has only one solution for the two cases as follows. 

Case 1: $\forall x \in (0,\lambda_1)$, according to the definition of concave function,  $f_1(x)>f_3(x)$. So $f_1 (\lambda)$ and $f_3(\lambda)$ have no intersection. $\forall \lambda \in (\lambda_1,1)$, as $f_1(\lambda)$ is a decreasing function, and $f_3(\lambda)$ is an increasing function, so in $(\lambda_1,1)$, $f_1 (\lambda)$ and $f_3(\lambda)$ also have no intersection. In this case $\lambda_1\geq\lambda_2$ and the optimal {value} $\lambda^*=\lambda_1$.

Case 2: $\forall x \in (0,\lambda_1)$, similar to Case 1, $f_1(x)>f_3(x)$ also holds. So $f_1 (\lambda)$ and $f_3(\lambda)$ have no intersection. $\forall \lambda \in(\lambda_1,1)$, according to the first-order condition of concave function, $f_3(\lambda)$ is located below its tangent line $f_4(\lambda)$ at the intersection point. That is, $f_4(\lambda)>f_3(\lambda)$. Obviously, $\forall \lambda \in (\lambda_1,1)$, $f_1(\lambda)>f_4(\lambda)$, so in $(\lambda_1,1)$, $f_1(\lambda)>f_3(\lambda)$, and $f_1 (\lambda)$ and $f_3(\lambda)$ have no intersection.  In this case $\lambda_1<\lambda_2$, the optimal {value} $\lambda^*=\lambda_2$.

In summary of above two cases, the optimal {value} $\lambda^*$ is computed as $\lambda^*=\max\{\lambda_1,\lambda_2\}.$ Now, we derive the expressions of $\lambda_1$ and $\lambda_2$, i.e., (\ref{lmd1}) and (\ref{lmd2}), by using Lambert-W
function $\mathcal{W}(x)$ as follows.

Let
$
t=\frac{\lambda}{1-\lambda}.
$
(\ref{f1eqf2}) is transformed into $
at=b+\log(1+ct)$.
Define $-z=at+a/c$, we have that $ze^z=-\frac{a}{c}e^{-b-a/c}$. So
$z=\mathcal{W}(-\frac{a}{c}e^{-b-a/c})$ and
\begin{equation}\label{tW}
t=-\frac{\mathcal{W}(-\frac{a}{c}e^{-b-\frac{a}{c}})}{a}-\frac{1}{c}.
\end{equation}
With $t=\frac{\lambda}{1-\lambda}$, the solution of (\ref{f1eqf2}) can be derived, which is
\begin{equation}
\lambda_1=\frac{-\frac{\mathcal{W}(-\frac{a}{c}e^{-b-\frac{a}{c}})}{a}-\frac{1}{c}}{1-\frac{\mathcal{W}(-\frac{a}{c}e^{-b-\frac{a}{c}})}{a}-\frac{1}{c}}.
\end{equation}

Then we derive the stagnation point of $f_1(\lambda)$ by setting its
first derivative equal to 0, i.e., $f_1'(\lambda)=0$.
Let
$s=1+\frac{c\lambda}{1-\lambda}$.
With $f_1'(\lambda)=0$, we get
$
\log(s)=\frac{c-1}{s}+1.
$
Let $\log(s)-1=u$. Then $e^uu=\frac{c-1}{e}$. Thus,
$u=\mathcal{W}(\frac{c-1}{e})$.
As a result,
\begin{equation}\label{s2}
s=e^{\mathcal{W}(\frac{c-1}{e})+1}.
\end{equation}
Therefore, with $s=1+\frac{c\lambda}{1-\lambda}$, the solution of $f_1'(\lambda)=0$, i.e., $\lambda_2$, is given by
\begin{equation}
\lambda_2=\frac{e^{\mathcal{W}(\frac{c-1}{e})+1}-1}{e^{\mathcal{W}(\frac{c-1}{e})+1}+c-1}.
\end{equation}
\end{IEEEproof}

\section{{Proof of Theorem 2}}
\begin{IEEEproof}
With a similar proof of Theorem 1 in Appendix A, it can be proved that there is one and only one solution associated with variable $\lambda$ in the interval (0,1) for the equation (\ref{f1eqf3}). Thus, we omit the related part here. Now, we only need to derive prove the expressions of $\lambda_1$ and $\lambda_2$ in (\ref{optimallambdaEC1}) and (\ref{optimallambdaEC2}).

Let $t=\frac{\lambda}{1-\lambda}$, (\ref{f1eqf3}) is transformed {into}
$at=\log(1+m+ct)$.
We have that
$
t=-\frac{\mathcal{W}(-\frac{a}{c}e^{-\frac{a(1+m)}{c}})}{a}-\frac{1+m}{c}.
$
Then we can obtain the solution of (\ref{f1eqf3}), which is
\begin{equation}
\lambda_1=\frac{-\frac{\mathcal{W}(-\frac{a}{c}e^{-\frac{a(1+m)}{c}})}{a}-\frac{1+m}{c}}{1-\frac{\mathcal{W}(-\frac{a}{c}e^{-\frac{a(1+m)}{c}})}{a}-\frac{1+m}{c}}.
\end{equation}

Moreover, the first derivative of $f_2(\lambda)$ defined in Lemma 3 with respect to $\lambda$ is
$
f_2'(\lambda)=-\log(1+m+\frac{c\lambda}{1-\lambda})+\frac{(c-(1+m))(1-\lambda)}{(1+m)(1-\lambda)+c\lambda}+b+1.
$
By letting the first derivative
\begin{equation}\label{dr2}
f_2'(\lambda)=0.
\end{equation}
and defining
$
w=1+m+\frac{c\lambda}{1-\lambda},
$
one can transform (\ref{dr2}) {into} $\log(w)=\frac{c-(1+m)}{w}+b+1.$
Let $\log(w)-(b+1)=v$. Then, $e^vv=\frac{c-(1+m)}{e^{(1+b)}}$. Thus,
$
v=\mathcal{W}(\frac{c-(1+m)}{e^{(1+b)}}).
$
As a result, we obtain $w=e^{v+b+1}$ and with $w=1+m+\frac{c\lambda}{1-\lambda}$ and $v=\mathcal{W}(\frac{c-(1+m)}{e^{(1+b)}})$, the solution of (\ref{dr2}), i.e., $\lambda$, is given by
\begin{equation}
\lambda_2=\frac{e^{\mathcal{W}(\frac{c-(1+m)}{e^{(1+b)}})+b+1}-(1+m)}{e^{\mathcal{W}(\frac{c-(1+m)}{e^{(1+b)}})+b+1}+c-(1+m)}.
\end{equation}

\end{IEEEproof}

\section{The selection of the branches of $\mathcal{W}(x)$}
\subsection{The Selection of the Branches of $\mathcal{W}(x)$ in (\ref{lmd1}) and (\ref{optimallambdaEC1})}

First we consider (\ref{lmd1}). According to $0<\lambda<1$ and $t=\frac{\lambda}{1-\lambda}$, we have that $t>0$. From (\ref{tW}), we obtain that
$
-\frac{\mathcal{W}(-\frac{a}{c}e^{-b-\frac{a}{c}})}{a}-\frac{1}{c}>0.
$
Let $K_1=-\frac{a}{c}$. Then, this {inequality} can be rewritten as
\begin{equation}\label{K1}
\mathcal{W}(K_1e^{-b+K_1})<K_1.
\end{equation}
Let $K_1e^{K_1}=K_2$, then $K_1=\mathcal{W}(K_2)$. Then, (\ref{K1})
can be rewritten as
$
\mathcal{W}(K_2e^{-b})<\mathcal{W}(K_2).
$
Due to $K_2<0$, $b>0$ and $0<e^{-b}<1$, we can derive
$K_2e^{-b}>K_2$. This indicates that $\mathcal{W}(x)$ is a
decreasing function. So in (\ref{lmd1}), the function
$\mathcal{W}(x)$ should select the branch $\mathcal{W}_{-1}(x)$.
The similar analysis above can be applied to (\ref{optimallambdaEC1}), for which $\mathcal{W}(x)$ should also select the branch $\mathcal{W}_{-1}(x)$.

\subsection{The Selection of the Branches of $\mathcal{W}(x)$ in (\ref{lmd2})  and (\ref{optimallambdaEC2})}
First we consider (\ref{lmd2}).
According to $0<\lambda<1$, $c>0$ and $s=1+\frac{c\lambda}{1-\lambda}$, we have that $s>1$. From (\ref{s2}), we obtain
$
s=e^{\mathcal{W}(\frac{c-1}{e})+1}>1.
$
This indicates that $\mathcal{W}(\frac{c-1}{e})>-1$, so in
(\ref{lmd2}), the function $\mathcal{W}(x)$ should select the branch
$\mathcal{W}_{0}(x)$.
For (\ref{optimallambdaEC2}), by adopting the similar method and $b=\log(1+m)$, $\mathcal{W}(x)$ should also select the branch $\mathcal{W}_{0}(x)$.

\subsection{The Selection of the Branches of $\mathcal{W}(x)$ in (\ref{alpha1}) and (\ref{TSalpha})}
First we consider (\ref{alpha1}).
Let $M_1=\frac{C_\textsf{SR}^\textrm{(TS)}}{\alpha_3}$, $M_2=-\frac{C_\textsf{SD}}{\alpha_3}$, $M_3=-\frac{c}{\alpha_3}$ and $M_4=\frac{c(1-\alpha_3)}{\alpha_3}+1$. Then, (\ref{alpha1}) can be expressed as
\begin{equation}\label{a1}
\alpha_1=-\frac{\mathcal{W}(-\frac{M_1}{M_3}e^{M_2-\frac{M_1M_4}{M_3}})}{M_1}-\frac{M_4}{M_3}.
\end{equation}
Let $N_1=-\frac{M_1}{M_3}$. Due to $\alpha_1>0$, from (\ref{a1}), we can obtain that
\begin{equation}\label{N1}
\mathcal{W}(N_1e^{M_2+dN_1})<dN_1.
\end{equation}
Let $N_2=M_4N_1e^{M_4N_1}$, then $M_4N_1=\mathcal{W}(N_2)$.
(\ref{N1}) can be transformed to
$$
\mathcal{W}\left(\frac{e^{M_2}}{M_4}N_2\right)<\mathcal{W}(N_2).
$$
Due to $M_2<0$ and $M_4>1$, we have that
$0<\frac{e^{M_2}}{M_4}<1$. Then, one can derive that
$\frac{e^{M_2}}{M_4}N_2<N_2$. This indicates that $\mathcal{W}(x)$
is a increasing function. So in (\ref{alpha1}), the function
$\mathcal{W}(x)$ should select the branch $\mathcal{W}_{0}(x)$.
For (\ref{TSalpha}), the similar method can be applied, {and thus} $\mathcal{W}(x)$ should also select the branch $\mathcal{W}_{0}(x)$.







%

\end{document}